\documentclass[journal,twocolumn,10pt]{IEEEtran}

\usepackage{graphicx}% Include figure files
\usepackage{dcolumn}% Align table columns on decimal point
\usepackage{bm}% bold math

\usepackage[ruled,vlined]{algorithm2e}
\usepackage{amsmath, amssymb}
\usepackage{booktabs}
\usepackage{multirow}
\usepackage{xcolor}
\usepackage{colortbl}
\usepackage{comment}
\usepackage{physics}
\usepackage{subfigure}
\usepackage{amsmath}
\usepackage{cite}
\usepackage{kotex}

\usepackage{soulutf8}
\RequirePackage[normalem]{ulem} %DIF PREAMBLE
\RequirePackage{color}\definecolor{RED}{rgb}{1,0,0}\definecolor{BLUE}{rgb}{0,0,1} %DIF PREAMBLE

\def\BibTeX{{\rm B\kern-.05em{\sc i\kern-.025em b}\kern-.08em
    T\kern-.1667em\lower.7ex\hbox{E}\kern-.125emX}}

\begin{document}

\title{Evolutionary BP+OSD Decoding for Low-Latency Quantum Error Correction}

\author{Hee-Youl Kwak, \IEEEmembership{Member,~IEEE}, 
Seong-Joon Park,~Hyunwoo Jung, Jeongseok Ha, \IEEEmembership{Senior Member,~IEEE}, and Jae-Won Kim, \IEEEmembership{Member,~IEEE}

\thanks{H.-Y. Kwak is with the Department of Electrical, Electronic and
Computer Engineering, University of Ulsan, Ulsan 44610, South Korea (e-mail: ghy1228@gmail.com).}
\thanks{S.-J. Park was with the Department of Electrical Engineering at Pohang University of Science and Technology (POSTECH), Pohang, Gyeongbuk 37673, South Korea  (e-mail: joonpark2247@gmail.com).}
\thanks{H. Jung and J. Ha are with the School of Electrical Engineering, Korea Advanced Institute of Science and Technology (KAIST), Daejeon 34141, South Korea (e-mail: destinylz@kaist.ac.kr, jsha@kaist.edu).}
\thanks{J.-W. Kim is with the Department of Electronic Engineering, Gyeongsang National University, Jinju 52828, South Korea  (e-mail: jaewon07.kim@gnu.ac.kr).}

\thanks{(Corresponding author: J.-W. Kim)}
}

\maketitle
\begin{abstract}
Quantum error correction (QEC) for fault-tolerant quantum computing requires a balanced decoding solution that offers high performance, low complexity, and low latency. However, the de facto standard—belief propagation (BP) combined with ordered statistics decoding (OSD)—suffers from excessive iterations in the BP stage and high complexity in the OSD stage. To address these challenges, we propose an evolutionary BP (EBP) decoder optimized via a differential evolution (DE) algorithm. By leveraging the gradient-free nature of DE, we enable end-to-end optimization of the EBP+OSD structure to maximize overall performance. In addition, a multi-objective selection rule is introduced to suppress frequent OSD activation, significantly reducing complexity overhead. Simulation results on surface codes and quantum low-density parity-check (QLDPC) codes demonstrate that EBP+OSD simultaneously achieves both superior decoding performance and substantially lower complexity compared to conventional BP+OSD, particularly in stringent low-latency regimes.
\end{abstract}

\maketitle

\begin{IEEEkeywords}
belief propagation (BP), differential evolution (DE), ordered statistics decoding (OSD), quantum error correction (QEC), quantum low-density parity-check (QLDPC) codes, surface codes
\end{IEEEkeywords}

\section{Introduction}
\label{sec_intro}
Quantum computers promise computational capabilities far exceeding classical systems in diverse fields such as integer factorization, search algorithms, optimization, and quantum chemistry~\cite{Shor1999,Grover1996, AspuruGuzik2005, GidneyEkera2021, Arute2019}.
However, the physical realization of quantum computers faces a fundamental challenge due to high error rates arising from noisy hardware and environmental decoherence.
Quantum error correction (QEC) addresses this limitation by encoding logical quantum information into multiple physical qubits~\cite{Shor1995, Steane1996, Calderbank1996, Gottesman1996}. 
Recent experimental breakthroughs have demonstrated substantial suppression of logical error rates (LERs) as the code distance increases, marking steady progress toward practical fault-tolerant quantum computation~\cite{Bluvstein2024, Bravyi2024, Acharya2023, GoogleQuantumAI2024}.

A key requirement for practical QEC is achieving reliable decoding within the coherence time of the quantum system, which demands extremely low latency and low computational complexity~\cite{Acharya2023, GoogleQuantumAI2024, Terhal2015, Forlivesi2025Bubble}. From a complexity standpoint, the belief propagation (BP) decoder \cite{Gallager1962} is a preferred candidate, especially for sparse codes such as the surface code~\cite{Kitaev2003, Dennis2002, BravyiKitaev1998, Fowler2012} and quantum low-density parity-check (QLDPC) codes~\cite{Bravyi2024, MacKay2004, PanteleevKalachev2021IT, PanteleevKalachev2021}, due to its linear complexity. However, its performance is severely limited by quantum degeneracy and short cycles inherent in quantum codes~\cite{Poulin2008, Babar2015, LaiKuo2021}. To mitigate this limitation, hybrid schemes that combine BP with ordered statistics decoding (OSD) have been proposed~\cite{PanteleevKalachev2021, RWBC2020}. While BP+OSD provides significantly improved decoding performance, the BP stage typically requires many iterations, resulting in high latency, and the OSD stage introduces substantial complexity overhead.

Machine learning-based decoders, including feed-forward networks (FFNs)~\cite{Varsamopoulos2017}, convolutional neural networks (CNNs)~\cite{ChamberlandRonagh2018, JungAliHa2024}, and transformers~\cite{Bausch2024}, have also been investigated. However, their high complexity limits scalability to large-scale codes. In contrast, the neural BP (NBP) decoder augments the BP decoder with trainable weights while preserving linear complexity~\cite{Nachmani2018, LiuPoulin2019, Miao2025}. Despite this advantage, its performance still lags behind that of the BP+OSD decoder. 
Although NBP can be combined with OSD, the non-differentiable operations of the OSD stage prevent end-to-end optimization by gradient descent. Consequently, the weights of NBP cannot be trained to work synergistically with the OSD stage.

In this paper, we propose an evolutionary BP (EBP) decoder. Similar to the NBP decoder, the EBP decoder employs trainable weights, but optimizes them using a differential evolution (DE) algorithm~\cite{StornPrice1997, DasSuganthan2011}. Unlike gradient-based methods, DE can handle non-differentiable operations, enabling end-to-end optimization of the overall EBP+OSD structure. Furthermore, we introduce a multi-objective selection rule to maximize decoding performance while simultaneously suppressing activation of the high-cost OSD stage. 

One limitation of DE is its degraded efficiency when optimizing a large number of weights. To address this issue, we introduce a weight-sharing method based on edge indices, which significantly reduces the number of trainable weights and improves optimization efficiency.
This approach also enables weight reuse across codes within the same class, thereby eliminating the need for time-consuming optimization of each individual code.

Simulation results show that the proposed EBP+OSD decoder outperforms the conventional BP+OSD decoder for both surface codes and QLDPC codes in terms of LER and threshold performance, while achieving lower decoding complexity through a substantial reduction in OSD activations. Notably, these improvements are obtained under strict low-latency constraints, using only 5 BP iterations—far fewer than those used in prior studies (BP+OSD~\cite{PanteleevKalachev2021}: 32 iterations; BP with topological blocking decoder (TBD)~\cite{JungHa2025}: 20 iterations; BP with memory~\cite{KuoLai2022}: over
150 iterations; relay BP~\cite{Muller2025Relay}: over 80 iterations; restart BP~\cite{Valentini2026}: over 50 iterations).
In summary, the proposed EBP+OSD decoder surpasses the conventional BP+OSD decoder across multiple metrics, including decoding performance, computational complexity, and suitability for low-latency operation, making it a highly compelling candidate for practical QEC systems.

The rest of the paper is organized as follows. Section~\ref{sec_pre} reviews the relevant preliminaries, including stabilizer codes, BP decoding, and the OSD algorithm. Section~\ref{sec_EBP} presents the DE-based optimization of the proposed EBP decoder, along with the weight sharing method. Section~\ref{sec_Results} compares the EBP+OSD decoder with conventional decoders in terms of decoding performance and computational complexity. Finally, Section~\ref{sec_conclusion} concludes the paper.

\section{Preliminaries}
\label{sec_pre}
\subsection{Quantum Stabilizer Codes}
Unlike classical error correction, QEC faces fundamental challenges, primarily the no-cloning theorem \cite{Wootters1982}, which prohibits the duplication of quantum states for redundancy, and the state collapse that occurs upon direct qubit measurement. Stabilizer codes overcome these issues by distributing logical information across multiple entangled physical qubits and extracting error syndromes through stabilizer measurements without disturbing the encoded logical state.

An $[[n,k,d]]$ stabilizer code encodes $k$ logical qubits into an entangled state $\ket\psi$ of $n$ physical (data) qubits \cite{Gottesman1996}. The code is defined by a stabilizer group $\mathcal{S}$, which is an Abelian subgroup of the $n$-qubit Pauli group $\mathcal{P}^{\otimes n}$, where $\mathcal{P} = \{I, X, Z, Y\}$. The group $\mathcal{S}$ is generated by $m=n-k$ independent and commuting generators $\{S_i\}_{i=1}^m$. This group can be represented by an $m \times n$ stabilizer matrix $\mathbf{S} \in \{I, X, Y, Z\}^{m \times n}$, where each row specifies the Pauli operators associated with $S_i$. Formally, the codespace $\mathcal{C}$ is defined as the common $+1$ eigenspace of all stabilizers in $\mathcal{S}$:
\begin{equation}
\mathcal{C} = \{ |\psi\rangle \mid S\ket\psi = (+1)\ket\psi, \forall S \in \mathcal{S} \}.
\end{equation}
The logical state $\ket\psi$ remains invariant under the action of any stabilizer. Logical operators $L \in \mathcal{L}$ are Pauli operators that commute with all stabilizers in $\mathcal{S}$ but are not elements of $\mathcal{S}$. For each logical qubit, there exist corresponding logical operators that map one logical state to another within the codespace $\mathcal{C}$. The minimum distance $d$ is defined as the minimum weight of any operator in $\mathcal{L}$.

\begin{figure}[t]
\centering
\subfigure[]{\includegraphics[scale=0.36]{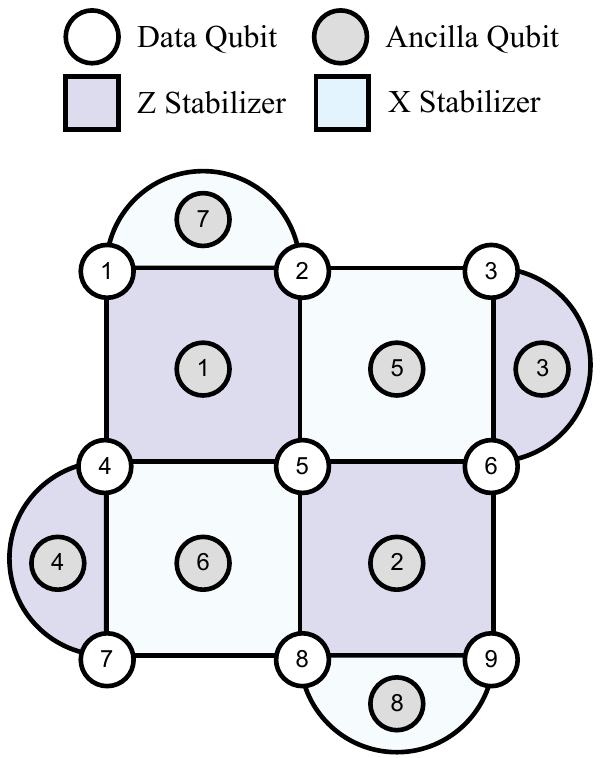}}
\subfigure[]{\includegraphics[scale=0.41]{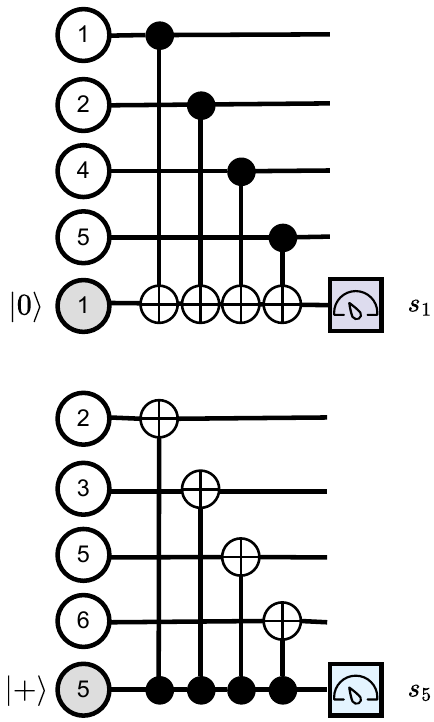}}
\subfigure[]{\includegraphics[scale=0.51]{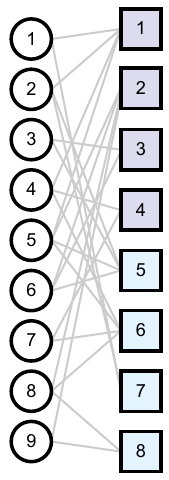}}
\caption{(a) Topological lattice of a $[[9,1,3]]$ surface code, where $n=9$ data qubits are positioned at the vertices and $m=8$ stabilizers are associated with the plaquettes. 
(b) Syndrome extraction circuits for $Z$-type and $X$-type stabilizers, where ancilla qubits are measured to obtain syndrome bits. 
(c) The corresponding Tanner graph representation, where the measured syndrome vector $\mathbf{s}$ is mapped onto CNs for the BP-based decoding.}
\label{Fig:Surface}
\end{figure}

The state $\ket\psi$ of an $[[n,k,d]]$ code resides in a $2^k$-dimensional subspace of a $2^n$-dimensional Hilbert space. A physical error $\mathcal{E} \in \mathcal{P}^{\otimes n}$ maps the state $\ket\psi$ to $\mathcal{E}\ket\psi$. To identify such errors, the syndrome vector $\mathbf{s}=(s_1, \ldots, s_m) \in \mathbb{F}_2^m$ is obtained via syndrome extraction circuits, where each syndrome bit $s_i$ is measured using an ancilla qubit associated with the stabilizer $S_i$. Mathematically, $s_i$ is represented by the symplectic inner product $s_i = \langle \mathcal{E}, S_i \rangle$, which equals $0$ if $\mathcal{E}$ and $S_i$ commute and $1$ if they anticommute. If all syndrome bits are zero, the state $\mathcal{E}\ket\psi$ remains within the codespace $\mathcal{C}$. Specifically, if the error $\mathcal{E}$ is an element of $\mathcal{S}$, the logical state remains unchanged; however, if $\mathcal{E}\in{\mathcal L}$, a logical error occurs despite the zero syndrome. Conversely, a non-zero syndrome bit indicates that the state has been projected into an error subspace orthogonal to $\mathcal{C}$. The task of the decoder is to estimate the most probable error given the observed syndrome vector $\mathbf{s}$.

Fig.~\ref{Fig:Surface} illustrates the overall QEC process using a $[[d^2,1,d]]$ rotated surface code with $d=3$. First, the lattice representation in Fig.~\ref{Fig:Surface}(a) shows the arrangement of $n=9$ data qubits at the vertices, while $m=8$ ancilla qubits are located at the plaquettes. Dark purple plaquettes represent $Z$-type stabilizers for detecting $X$ errors, whereas light blue ones represent $X$-type stabilizers for detecting $Z$ errors. The syndrome extraction circuits in Fig.~\ref{Fig:Surface}(b) illustrate how ancilla qubits interact with neighboring data qubits to generate syndrome bits, such as $s_1$ and $s_5$. The Tanner graph representation in Fig.~\ref{Fig:Surface}(c) is used for BP decoding, which is described in detail in the next subsection.

\subsection{Quantum BP Decoding}
The decoding objective is to estimate the most probable error operator $\hat{\mathcal{E}}$ such that $\mathcal{E}\hat{\mathcal{E}} \in \mathcal{S}$ given the observed syndrome $\mathbf{s}$. 
The BP algorithm is an iterative message-passing algorithm that updates log-likelihood ratio (LLR) messages over the Tanner graph. This graph consists of $n$ variable nodes (VNs), representing physical qubits, and $m$ check nodes (CNs), representing stabilizers. The graph structure is defined by the stabilizer matrix $\mathbf{S}$, where $S_{i,j}$ denotes the Pauli operator acting on VN $v_j$ for CN $c_i$, and an edge exists between them if $S_{i,j}$ is non-identity.

For quantum codes, BP can be implemented as either binary ${\rm BP}_2$ \cite{Babar2015}, which treats $X$ and $Z$ errors independently, or as quaternary ${\rm BP}_4$ \cite{LaiKuo2021}, which accounts for the correlation between $X$ and $Z$ components.
NBP extends conventional BP by introducing trainable weights into the message update rules \cite{Nachmani2018}. In NBP, the Tanner graph is unrolled into a deep neural network, and the weights are optimized via gradient descent. NBP for quantum codes also has two variants: ${\rm NBP}_2$ \cite{LiuPoulin2019} and ${\rm NBP}_4$ \cite{Miao2025}. In this work, we focus on ${\rm BP}_4$ and ${\rm NBP}_4$, which are hereafter referred to as BP and NBP, respectively.

Let $\mathcal{E}_j\in \mathcal{P}$ denote the $j$-th element of the physical error $\mathcal{E}$. Under a depolarizing noise model with error probability $p$, an error $\zeta \in \{X, Z, Y\}$ occurs at each VN with probability $p/3$. For each VN $v_j$ with error $\zeta$ and its connected CN $c_i$, the VN messages ${m}_{j \to i,\zeta}^{(\ell)}$ at iteration $\ell$ are initialized as ${m}_{j \to i,\zeta}^{(0)} = L_{j,\zeta}$, where
\begin{equation}
L_{j,\zeta} = \ln \left( \frac{P(\mathcal{E}_j=I)}{P(\mathcal{E}_j=\zeta)} \right) = \ln \left( \frac{1-p}{p/3} \right).
\end{equation}
Since the exact channel parameter $p$ is often unknown and has limited impact on decoding performance \cite{Miao2025}, we fix $p=0.1$ for the initialization.

After the initialization, the decoder performs iterative updates from $\ell=1$ to $\overline{\ell}$. First, the belief quantization converts the VN message vector $\left({m}_{j \to i,X}^{(\ell-1)},{m}_{j \to i,Z}^{(\ell-1)},{m}_{j \to i,Y}^{(\ell-1)}\right)$ into the scalar message $\lambda_{j \to i}^{(\ell-1)}$, which represents the belief that the error $\mathcal{E}_j$ commutes with $S_{i,j}$:
\begin{equation}
\lambda_{j \to i}^{(\ell-1)} = \ln \left( \frac{1 + \exp\left(-m_{j \to i, S_{i,j}}^{(\ell-1)}\right)}{\exp\left(-m_{j \to i, \zeta_1}^{(\ell-1)}\right) + \exp\left(-m_{j \to i, \zeta_2}^{(\ell-1)}\right)} \right),
\label{Eq:Belief_Quantization}
\end{equation}
where $\{\zeta_1, \zeta_2\} = \{X, Z, Y\} \setminus \{S_{i,j}\}$. Using the min-sum rule, the CN message $\overline{m}_{i \to j}^{(\ell)}$ from $c_i$ to $v_j$ is updated as
\begin{equation}
\overline{m}_{i \to j}^{(\ell)} = (-1)^{s_i} \! \left( \prod_{j' \in \mathcal{N}_c(i) \setminus \{j\}}\!\!\!\! \text{sgn}(\lambda_{j' \to i}^{(\ell-1)}) \right) \!\! \min_{j' \in \mathcal{N}_c(i) \setminus \{j\}} \! \big|  \lambda_{j' \to i}^{(\ell-1)} \big|,
\label{Eq:Min}
\end{equation}
where $\mathcal{N}_c(i)$ denotes the set of VNs indices neighboring $c_i$. 
Subsequently, the VN message for each error type $\zeta$ is updated as
\begin{equation}
m_{j \to i, \zeta}^{(\ell)} = \overline{w}_{j}^{(\ell)} L_{j,\zeta} + \sum_{\substack{i' \in \mathcal{N}_v(j) \setminus \{i\} \\ \langle \zeta, S_{i',j} \rangle = 1}} {w}_{i' \to j}^{(\ell)} \overline{m}_{i' \to j}^{(\ell)},
\label{Eq:Weighted_Sum}
\end{equation}
where $\mathcal{N}_v(j)$ denotes the set of CN indices neighboring $v_j$, and the parameters $\overline{w}_{j}^{(\ell)}$ and ${w}_{i' \to j}^{(\ell)}$ are trainable VN and CN weights assigned to the initial LLR and CN messages, respectively. The weight set $\mathcal{W} = \{\overline{w}_{j}^{(\ell)}, {w}_{i' \to j}^{(\ell)}\}$ is optimized in the unrolled neural network. When all weights are set to one, the algorithm reduces to standard BP.

At each iteration, the posterior LLR for VN $v_j$ and error type $\zeta$ is computed as
\begin{equation}
m_{j, \zeta} = \overline{w}_{j}^{(\ell)} L_{j, \zeta} + \sum_{\substack{i' \in \mathcal{N}_v(j) \\ \langle \zeta, S_{i',j} \rangle = 1}} {w}_{i' \to j}^{(\ell)} \overline{m}_{i' \to j}^{(\ell)}.
\label{Eq:Decision}
\end{equation}
The estimated error $\hat{\mathcal{E}}_j$ is then determined by a hard decision:
\begin{equation}
\hat{\mathcal{E}}_j =
\begin{cases}
I, & \text{if } m_{j,\zeta} > 0, \forall \zeta \in \{X, Z, Y\} \\
\mathop{\mathrm{arg\,min}}\limits_{\zeta} ~m_{j,\zeta}, & \text{otherwise.}
\end{cases}
\end{equation}
From the estimated error $\hat{\mathcal{E}}$, the corresponding syndrome $\hat{\mathbf{s}}$ is computed. Decoding terminates if the syndrome $\hat{\mathbf{s}}$ of the estimated error $\hat{\mathcal{E}}$ matches the observed syndrome $\mathbf{s}$, or when the maximum number of iterations $\overline{\ell}$ is reached. If no syndrome-matching $\hat{\mathcal{E}}$ is found within $\overline{\ell}$ iterations, a flagged failure is declared. In cases where $\hat{\mathbf{s}} = \mathbf{s}$, decoding is successful if $\mathcal{E}\hat{\mathcal{E}} \in \mathcal{S}$. Conversely, an unflagged failure (logical error) occurs if $\mathcal{E}\hat{\mathcal{E}} \in \mathcal{L}$.  The overall decoding failure rate is referred to as the LER.

\subsection{OSD Algorithm for Post-Decoder}
Upon a flagged failure of the BP or NBP decoder, a post-decoder is activated to find a syndrome-matching error estimate. In this pre+post decoding structure, the OSD algorithm is widely used due to its robustness across various quantum codes. OSD processes $X$ and $Z$ errors independently using the binary parity-check matrices $\mathbf{H}_Z\in \mathbb{F}_2^{m_Z \times n}$ and $\mathbf{H}_X\in \mathbb{F}_2^{m_X \times n}$ derived from the stabilizer matrix $\mathbf{S}$, where $m_Z$ and $m_X$ denote the numbers of $Z$ and $X$-type stabilizers, respectively. Without loss of generality, we describe the OSD procedure for $X$ errors using $\mathbf{H}_Z$.

The OSD with order-$0$ algorithm consists of the following steps. First, the posterior LLRs $m_{j,\zeta}$ are converted into unreliability measures $u_{j,X}$:
\begin{equation}
u_{j,X} =p_{j,X} + p_{j,Y},
\label{Eq:OSD_u}
\end{equation}
where the probability $p_{j,\zeta}$ of each error type $\zeta$ at VN $v_j$ is given by
\begin{equation}
p_{j,\zeta} = \frac{\exp(-m_{j, \zeta})}{1 + \sum_{\zeta' \in \{X, Z, Y\}} \exp(-m_{j, \zeta'})}.
\label{Eq:OSD_p}
\end{equation}
A larger value of $u_{j,X}$ indicates a higher likelihood of an $X$-type error on the $j$-th qubit. Second, the columns of $\mathbf{H}_Z$ are permuted in descending order of $u_{j,X}$. Third, Gaussian elimination is applied to the permuted matrix and syndrome vector to identify a set of linearly independent columns, which form a basis. The error values for the non-basis positions are set to zero, and the error vector on the basis is obtained by solving the resulting linear system. Finally, the estimated error vector is re-ordered to its original indices to produce the $X$-error estimate $\hat{\mathcal{E}}_X$. By combining the estimated $X$ and $Z$ error components, the OSD stage produces a syndrome-matching output $\hat{\mathcal{E}}$ such that $\hat{\mathbf{s}} = \mathbf{s}$.

\begin{algorithm}[t]
\caption{DE for the EBP Decoder}
\label{Alg:Proposed}
\DontPrintSemicolon
\textbf{Initialization:} Randomly generate $\mathcal{W}_t$ for $1\le t\le T$

\For{$g = 1$ \KwTo $G$}{
  \tcc{Mutation}
  \For{$t = 1$ \KwTo $T$}{
    Select distinct $t_1, t_2, t_3 \in \{1, \dots, T\} \setminus \{t\}$\;
    \If{$g > G/2$}{
        $t_1$ is set to the best index
    }
    Generate a mutant instance \begin{equation*} \overline{\mathcal{W}}_t = \mathcal{W}_{t_1} + F(\mathcal{W}_{t_2} - \mathcal{W}_{t_3})
    \end{equation*}
  }

  \tcc{Crossover}
  \For{$t = 1$ \KwTo $T$}{
    Generate a trial instance $\hat{\mathcal{W}}_t$ where
    \begin{equation*}
    \hat{\mathcal{W}}_{t,i} = \begin{cases} \overline{\mathcal{W}}_{t,i}, & \text{with probability } p_c \\ \mathcal{W}_{t,i}, & \text{otherwise} \end{cases}\end{equation*}
  }

  \tcc{Selection}
  \For{$t = 1$ \KwTo $T$}{
    Run EBP+OSD with $\mathcal{W}_t$ and $\hat{\mathcal{W}}_t$ and compute \begin{equation*}
\eta = \frac{\text{LER}_{+}(\mathcal{W}_t) - \text{LER}_{+}(\hat{\mathcal{W}}_t)}{\text{LER}_{+}(\mathcal{W}_t)}.
\end{equation*}

    \textbf{if} $\eta > \epsilon$ \;\textbf{~~then} $\mathcal{W}_t \leftarrow \hat{\mathcal{W}}_t$\;
    \textbf{else if} $|\eta| \le \epsilon$ \textbf{and} $\text{FFR}_{\text{P}}(\hat{\mathcal{W}}_t) < \text{FFR}_{\text{P}}(\mathcal{W}_t)$ \;
    \textbf{~~then} $\mathcal{W}_t \leftarrow \hat{\mathcal{W}}_t$\;
  }
}
\textbf{Output:} $\mathcal{W}_{\mathrm{opt}} = \arg\min_{\mathcal{W}_t} \text{LER}_{+}(\mathcal{W}_t)$
\end{algorithm}

\section{Evolutionary BP}
\label{sec_EBP}
\subsection{Motivation and Problem Formulation}
In a pre+OSD decoding structure, the OSD post-decoder is activated only upon a flagged failure of the pre-decoder. Accordingly, the overall LER of the combined scheme, denoted by $\text{LER}_{+}$, is expressed as
\begin{equation}\text{LER}_{+} = \text{UFR}_{\text{P}} + \text{FFR}_{\text{P}} \times \text{UFR}_{\text{O}},\label{Eq:LER_plus}
\end{equation}
where $\text{UFR}_{\text{P}}$ and $\text{FFR}_{\text{P}}$ denote the unflagged failure rate (UFR) and flagged failure rate (FFR) of the pre-decoder, respectively, and $\text{UFR}_{\text{O}}$ represents the unflagged failure rate of the OSD post-decoder. Here, $\text{FFR}_{\text{P}}$ also corresponds to the activation probability of the OSD stage. Note that the LER of a standalone pre-decoder such as BP or NBP is simply given by $\text{LER}_\text{P}=\text{UFR}_{\text{P}} + \text{FFR}_{\text{P}}$ .

The standard BP decoder often suffers from a high $\text{FFR}_{\text{P}}$ for quantum codes and the subsequent OSD decoder can effectively resolve these flagged failures to achieve a competitive $\text{LER}_{+}$ of the BP+OSD decoder. However, the frequent activation of OSD—arising from a high $\text{FFR}_{\text{P}}$—incurs substantial computational overhead. In contrast, the NBP decoder mitigates flagged failures by optimizing its trainable weights. However, this optimization often leads to an unintended increase in $\text{UFR}_{\text{P}}$ \cite{LiuPoulin2019}. Since unflagged failures bypass the OSD stage, the opportunity for additional error correction is lost, which ultimately degrades the overall $\text{LER}_{+}$ of the NBP+OSD decoder.

In summary, an ideal pre-decoder for the combined decoding structure should simultaneously satisfy two objectives: (i) minimizing $\text{LER}_{+}$ by providing OSD-friendly inputs, and (ii) reducing the OSD activation probability $\text{FFR}_{\text{P}}$ to lower the overall decoding complexity. Unlike the BP decoder, whose structure
is fixed, the NBP decoder can adapt its trainable
weights to meet these objectives. However, designing a
loss function that explicitly encourages such behavior remains
challenging. Moreover, directly training
the NBP+OSD decoder to minimize $\text{LER}_{+}$ is infeasible
because the OSD decoder involves inherently non-differentiable operations.

\subsection{DE Algorithm for EBP}
We propose the EBP decoder as an end-to-end optimizable pre-decoder used in conjunction with the post-decoder. The EBP decoder optimizes the weight set $\mathcal{W}$ using the DE algorithm. DE offers a distinct advantage over gradient-descent-based neural networks: it can handle non-differentiable objective functions and allows for flexible multi-objective optimization. Analogous to NBP, which is named for its use of neural networks, EBP derives its name from the application of evolutionary algorithms for weight optimization.

\begin{figure*}[t]
\centering
\subfigure[]{
\includegraphics[width=0.7\textwidth]{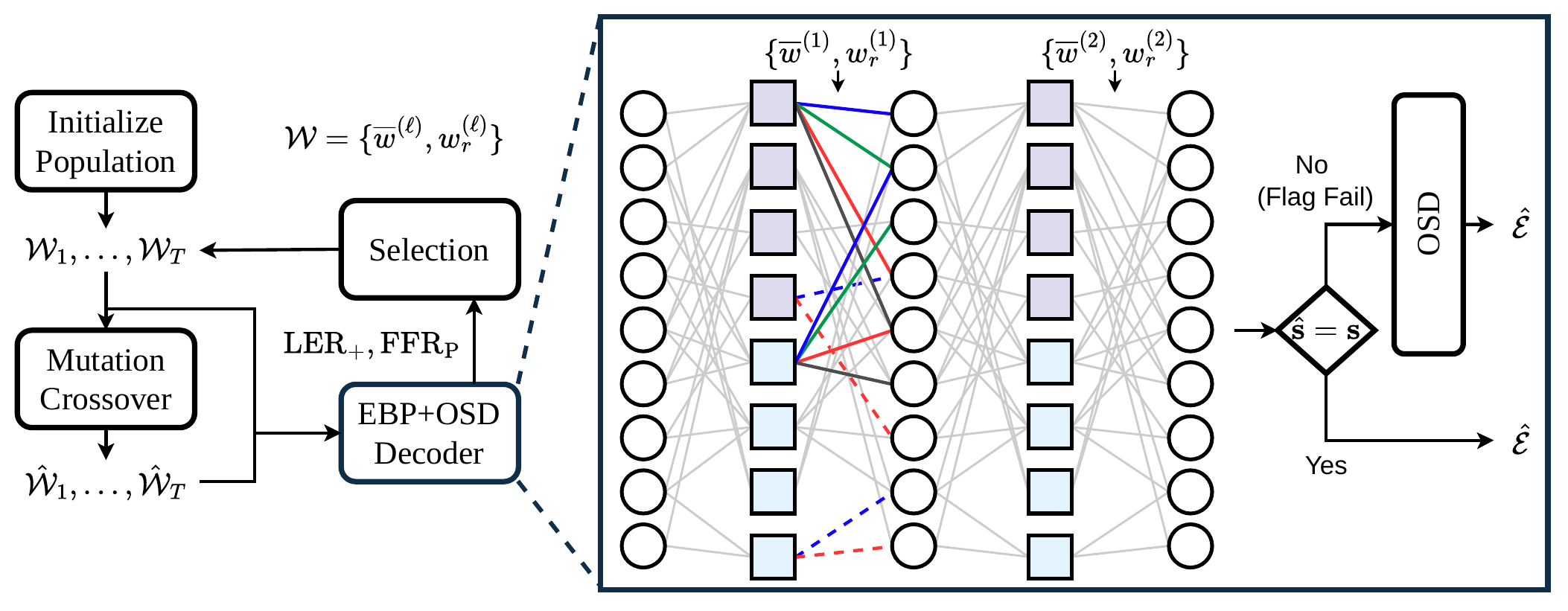}
}
\subfigure[]{
\includegraphics[width=0.26\textwidth]{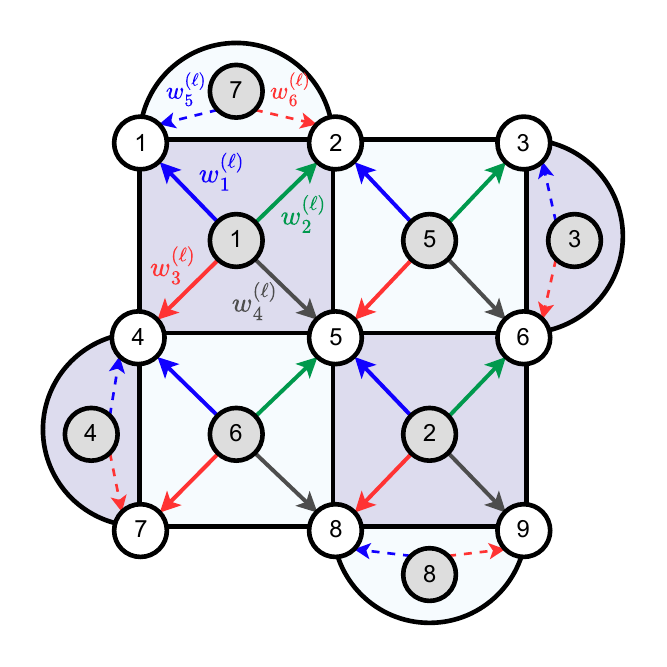}
}
\caption{(a) DE algorithm for optimizing the weight set $\mathcal{W}$ in the EBP decoder, where the objective directly targets the overall EBP+OSD performance rather than the standalone EBP performance, by incorporating the EBP+OSD decoder into the optimization loop. (b) The proposed weight-sharing method for the $d=3$ surface code, where edges are assigned identical weights based on their colors.}
\label{Fig:block_diagram}
\end{figure*}

The proposed DE algorithm for the EBP decoder is summarized in Algorithm~\ref{Alg:Proposed} and illustrated in Fig.~\ref{Fig:block_diagram}(a). 
The optimization variable is the weight set $\mathcal{W}$.
Initially, a population of $T$ candidate weight sets $\{\mathcal{W}_t\}_{t=1}^T$ is generated, where each element of $\mathcal{W}_t$ is randomly initialized in the range $[0,2]$. Each individual then evolves over $G$ generations through mutation, crossover, and selection.

\begin{figure*}[t]
\centering
\subfigure[]{\includegraphics[width=0.38\textwidth]{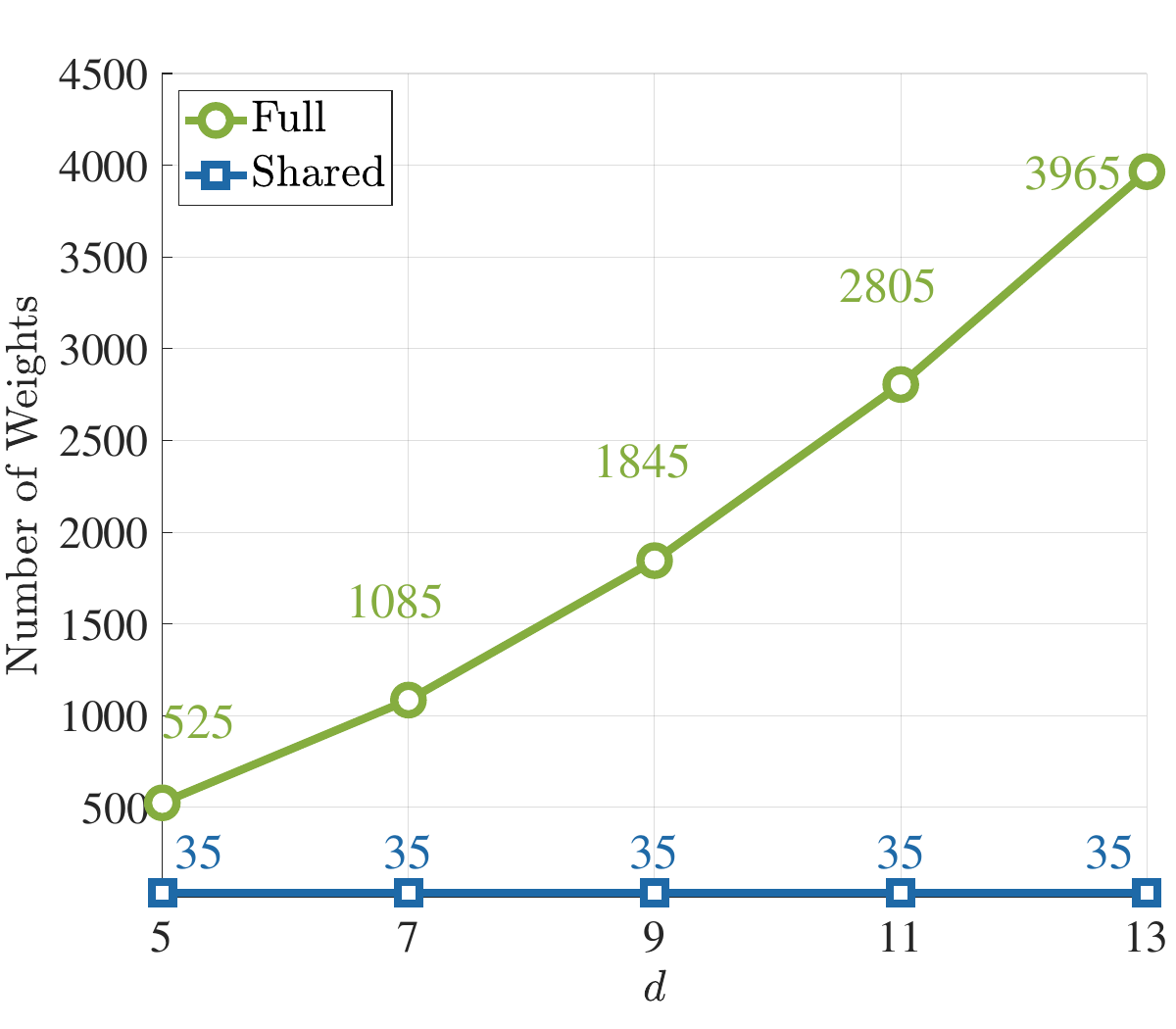}}
\quad
\subfigure[]{\includegraphics[width=0.38\textwidth]{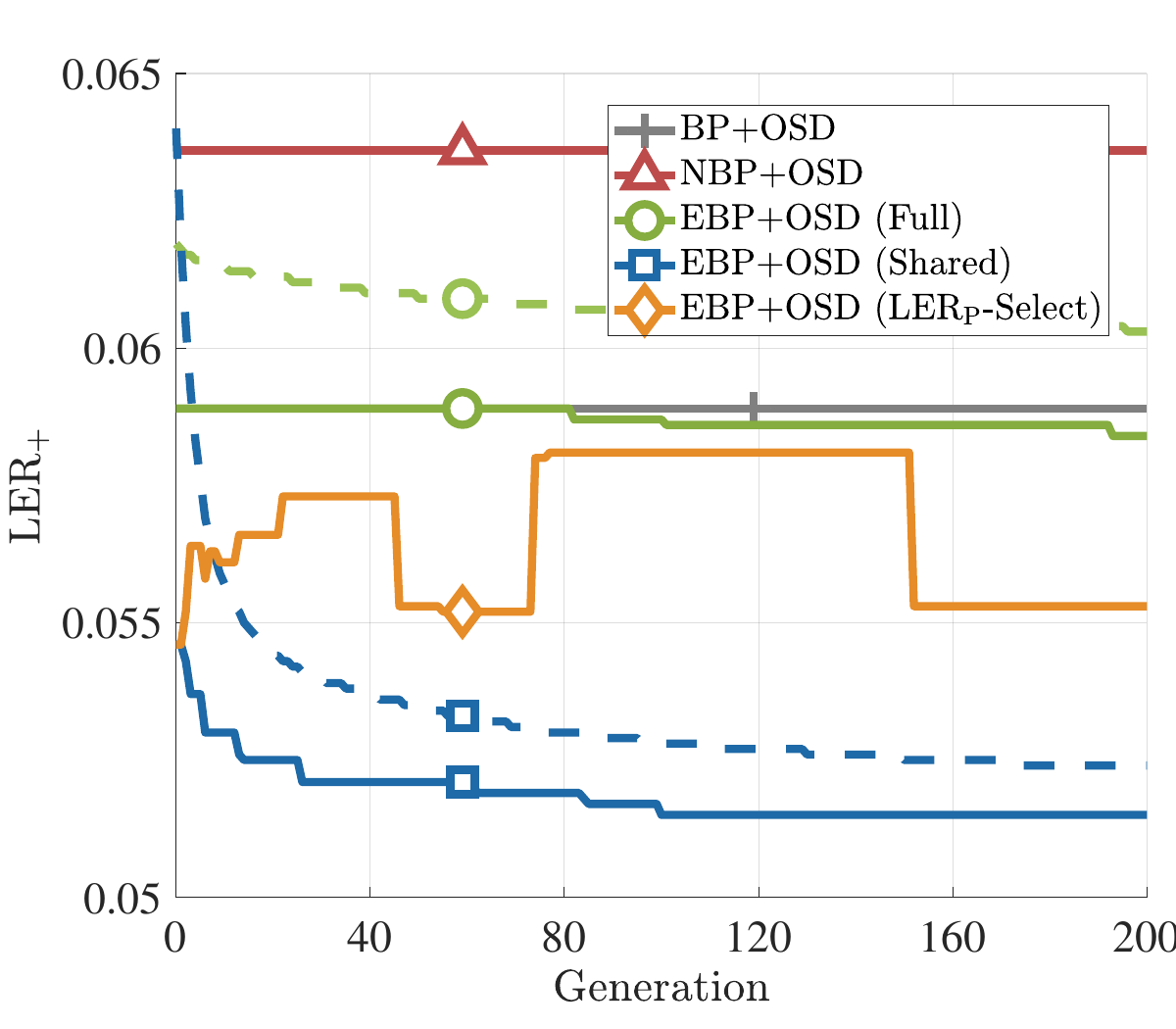}}
\caption{Effect of the proposed weight-sharing method and end-to-end optimization:
(a) The number of weights is significantly reduced and remains constant across different $d$ in surface codes.
(b) The proposed sharing method of the EBP+OSD decoder (Shared) accelerates DE convergence compared with the full parameterization case (Full) for the surface code with $d=7$, achieving a lower ${\rm LER}_{+}$ at the final generation and outperforming the baseline decoders BP+OSD and NBP+OSD. When the selection is instead driven by the standalone pre-decoder performance ${\rm LER}_{\rm P}$ (${\rm LER}_{\rm P}$-Select), the overall ${\rm LER}_{+}$ may even degrade.}
\label{Fig:Sharing}
\end{figure*}

In the mutation step, a mutant instance $\overline{\mathcal W}_t$ is generated from three individuals ${\mathcal W}_{t_1}$, ${\mathcal W}_{t_2}$, and ${\mathcal W}_{t_3}$ as
\begin{equation}
\overline{\mathcal{W}}_{t} = \mathcal{W}_{t_1} + F(\mathcal{W}_{t_2} - \mathcal{W}_{t_3}),
\end{equation}
where $F$ is a scaling factor and $t_1, t_2, t_3$ are initially selected as mutually distinct indices from $\{1, \dots, T\} \setminus \{t\}$. To balance exploration and convergence, we adopt an adaptive mutation strategy for selecting the index $t_1$. While $t_1$ remains a randomly chosen index (i.e., rand/1/bin strategy \cite{DasSuganthan2011}) during the first half of the generations ($g \le G/2$), it is reassigned to the index of the best-performing individual (i.e., best/1/bin strategy \cite{DasSuganthan2011}) for $g > G/2$ to accelerate convergence. Subsequently, a trial instance $\hat{\mathcal{W}}_t$ is formed by a crossover step, where the $i$-th element $\hat{\mathcal{W}}_{t,i}$ is sampled from either $\overline{\mathcal{W}}_{t,i}$ with probability $p_c$ or $\mathcal{W}_{t,i}$ otherwise:
\begin{equation}
\hat{\mathcal{W}}_{t,i} =
\begin{cases}
\overline{\mathcal{W}}_{t,i}, & \text{with probability } p_c, \\
\mathcal{W}_{t,i}, & \text{otherwise.}
\end{cases}
\end{equation}

In the selection step, we adopt the decoder-in-the-loop approach \cite{Elkelesh2019} to directly assess the performance of each weight set.
Both $\mathcal{W}_t$ and $\hat{\mathcal{W}}_t$ are evaluated using the EBP+OSD decoder through $10^5$ Monte Carlo trials under a depolarizing noise model with error probability $p_e$. Let $\text{LER}_{+}(\mathcal{W})$ and $\text{FFR}_{\text{P}}(\mathcal{W})$ denote $\text{LER}_{+}$ and $\text{FFR}_{\text{P}}$ given the weight set $\mathcal{W}$, respectively. To quantify performance improvement, we define the relative improvement ratio $\eta$ as
\begin{equation}
\eta = \frac{\text{LER}_{+}(\mathcal{W}_t) - \text{LER}_{+}(\hat{\mathcal{W}}_t)}{\text{LER}_{+}(\mathcal{W}_t)}.
\end{equation}
The trial instance $\hat{\mathcal{W}}_t$ replaces $\mathcal{W}_t$ for the next generation according to the following multi-objective selection rule with threshold $\epsilon$:
\begin{equation}\mathcal{W}_t \leftarrow
\begin{cases}\hat{\mathcal{W}}_t, & \text{if } \eta > \epsilon, \\
\hat{\mathcal{W}}_t, & \text{if } |\eta| \le \epsilon \text{ and } \text{FFR}_{\text{P}}(\hat{\mathcal{W}}_t) < \text{FFR}_{\text{P}}(\mathcal{W}_t), \\
\mathcal{W}_t, & \text{otherwise.}
\end{cases}
\end{equation}
This selection rule prioritizes individuals that achieve a significant improvement in $\text{LER}_{+}$. When the performance difference is marginal (within the $\epsilon$-indifference zone), the individual with lower computational complexity, reflected by a smaller $\text{FFR}_{\text{P}}$, is selected.

This evolutionary process is repeated for $G$ generations, and the individual with the best performance is selected as the final solution. In our implementation, the DE hyperparameters are set to $T = 500$, $G = 200$, $F = 0.5$, $p_c = 0.7$, $p_e = 0.1$, and $\epsilon=0.01$.
On an AMD EPYC 7763 processor, one decoding trial for the $d=7$ surface code requires approximately $2~\mu$s; thus, evaluating $10^5$ Monte Carlo trials requires about $200$ ms. With $T=500$ individuals and $G=200$ generations, it requires approximately $5.6$ hours. This cost is incurred only offline, and is further amortized because the learned shared weights can be reused across code distances; moreover, the Monte Carlo evaluations of population members can be executed in parallel.

\subsection{Edge-Indexed Weight-Sharing}
Although the DE algorithm is a robust tool for optimizing non-differentiable problems, its convergence typically degrades as the dimensionality of the search space increases. The total number of weights in the set $\mathcal{W} = \{\overline{w}_{j}^{(\ell)}, {w}_{i' \to j}^{(\ell)}\}$ is $\overline{\ell} \times (n + e)$, where $e$ denotes the number of edges in the Tanner graph. For instance, an $[[d^2, 1, d]]$ surface code requires $\overline{\ell} \times (d^2 + 4d(d+1))$ weights. For $d=11$ and $\overline{\ell}=5$, this amounts to $2{,}805$ weights, a scale that significantly degrades the efficiency of DE-based optimization.
To reduce the number of weights, we employ the weight sharing technique.
For VN weights, we adopt spatial sharing \cite{Dai2021}, where a single weight is applied to all VNs at each iteration, i.e., $\overline{w}_j^{(\ell)} = \overline{w}^{(\ell)}$.
For CN weights ${w}_{i \to j}^{(\ell)}$, however, simple spatial sharing can cause message trapping due to the symmetric structures inherent in quantum codes~\cite{RaveendranVasic2021}.

To address this issue, we introduce an edge-indexed sharing method, in which weights are shared among CNs of the same degree and assigned based on their local edge indices.
Consider a code with two CN degree types, $r_1$ and $r_2$, such as surface or QLDPC codes.
For example, the surface code with $d=3$ in Fig.~\ref{Fig:Surface} includes four CNs with $r_1 = 4$ and four CNs with $r_2 = 2$.
The edge-indexed sharing method defines a set of CN weights
$\{w_1^{(\ell)}, \ldots, w_{\overline{r}}^{(\ell)}\}$
for $\overline{r} = r_1 + r_2$.
The first $r_1$ weights are assigned to edges of degree-$r_1$ CNs, while the remaining $r_2$ weights are assigned to edges of degree-$r_2$ CNs.
Specifically, the $j$-th edge of a degree-$r_1$ CN uses $w_j^{(\ell)}$, whereas the $j$-th edge of a degree-$r_2$ CN uses $w_{r_1+j}^{(\ell)}$.
Accordingly, the weight set is reduced to
\begin{equation}
\mathcal{W} = \{ \overline{w}^{(\ell)}, w_r^{(\ell)} |1\le\ell\le \overline{\ell},1\le r \le \overline{r}\}.
\end{equation}
In Fig.~\ref{Fig:block_diagram}(b), for four degree-4 CNs, solid edges of the same color share the identical weight, while dashed edges connected to degree-2 CNs also share weights by colors, following the same principle.

Fig.~\ref{Fig:Sharing}(a) compares the numbers of weights in surface codes for $\overline{\ell}=5$ as a function of $d$, for both the full parameterized case (Full) and the proposed sharing method (Shared). The proposed sharing method significantly reduces the number of weights and the size of the associated search space. In addition, the number of weights without sharing increases quadratically with $d$, whereas the proposed method keeps it constant as ${\overline{\ell}\times (\overline{r}+1)}$ for all $d$.
This property allows a single optimized weight set to be reused across different code distances within the same code family, thereby substantially reducing the overall optimization time.

Fig.~\ref{Fig:Sharing}(b) shows the evolution of $\text{LER}_{+}$ over generations for the $d=7$ surface code at $p=0.1$ and $\overline{\ell}=5$. To focus on the effect of weight sharing, we adopt an LER-only selection rule ($\epsilon = 0$) and the rand/1/bin mutation strategy. We compare Full and Shared, by tracking both the population average (dotted lines) and best individual (solid lines). Without sharing, the population average fails to converge toward the best-performing individual. In contrast, with sharing, the average and best curves closely align, demonstrating that weight sharing accelerates DE convergence. Moreover, the EBP+OSD decoder with the proposed sharing method at the final generation significantly outperforms baseline decoders such as BP+OSD and NBP+OSD. In particular, NBP+OSD performs worse than BP+OSD, which indicates that optimizing the pre-decoder in isolation can be detrimental within the combined pre+OSD structure. To further illustrate this effect, we include a variant (${\rm LER}_{\rm P}$-Select) whose DE selection is driven by the standalone pre-decoder performance ${\rm LER}_{\rm P}$ instead of ${\rm LER}_{+}$. Although this variant can improve the standalone EBP performance, its overall ${\rm LER}_{+}$ varies erratically across generations, confirming that end-to-end optimization of the entire EBP+OSD structure is essential.

\begin{table}[t]
\centering
\small
\setlength{\tabcolsep}{2.5pt}
\caption{Performance analysis of EBP+OSD with various optimization configurations against BP+OSD and NBP+OSD}
\label{Table:Analysis}
\begin{tabular}{cccccccc}
\toprule
\multirow{2}{*}{Decoder} & \multirow{2}{*}{Shared} & \multirow{2}{*}{Multi.}  & \multirow{2}{*}{Adap.}  & ${\rm UFR}_{\rm P}$ & ${\rm FFR}_{\rm P}$ & ${\rm UFR}_{\rm O}$ & ${\rm LER}_{\rm +}$ \\
 &  & &  & $10^{-4}$ & $10^{-1}$ & $10^{-2}$ & $10^{-3}$\\
\midrule
EBP+OSD & X & X & X & $2.70$ & $1.46$ & $3.16$ & $4.89$ \\
EBP+OSD & O & X & X & $3.50$ & $1.26$ & $2.96$ & $4.08$ \\
EBP+OSD & O & O & X & $4.80$ & $0.95$ & $4.03$ & $4.31$ \\
\rowcolor{gray!10}
EBP+OSD & O & O & O & $5.60$ & $0.77$ & $4.52$ & $4.05$ \\
BP+OSD & - & - & - & $1.60$ & $3.74$ & $1.32$ & $5.10$ \\
NBP+OSD & - & - & - & $9.50$ & $0.70$ & $7.73$ & $6.39$ \\
\bottomrule
\end{tabular}
\end{table}

\begin{figure}[t]
\centering
\includegraphics[width=0.38\textwidth]{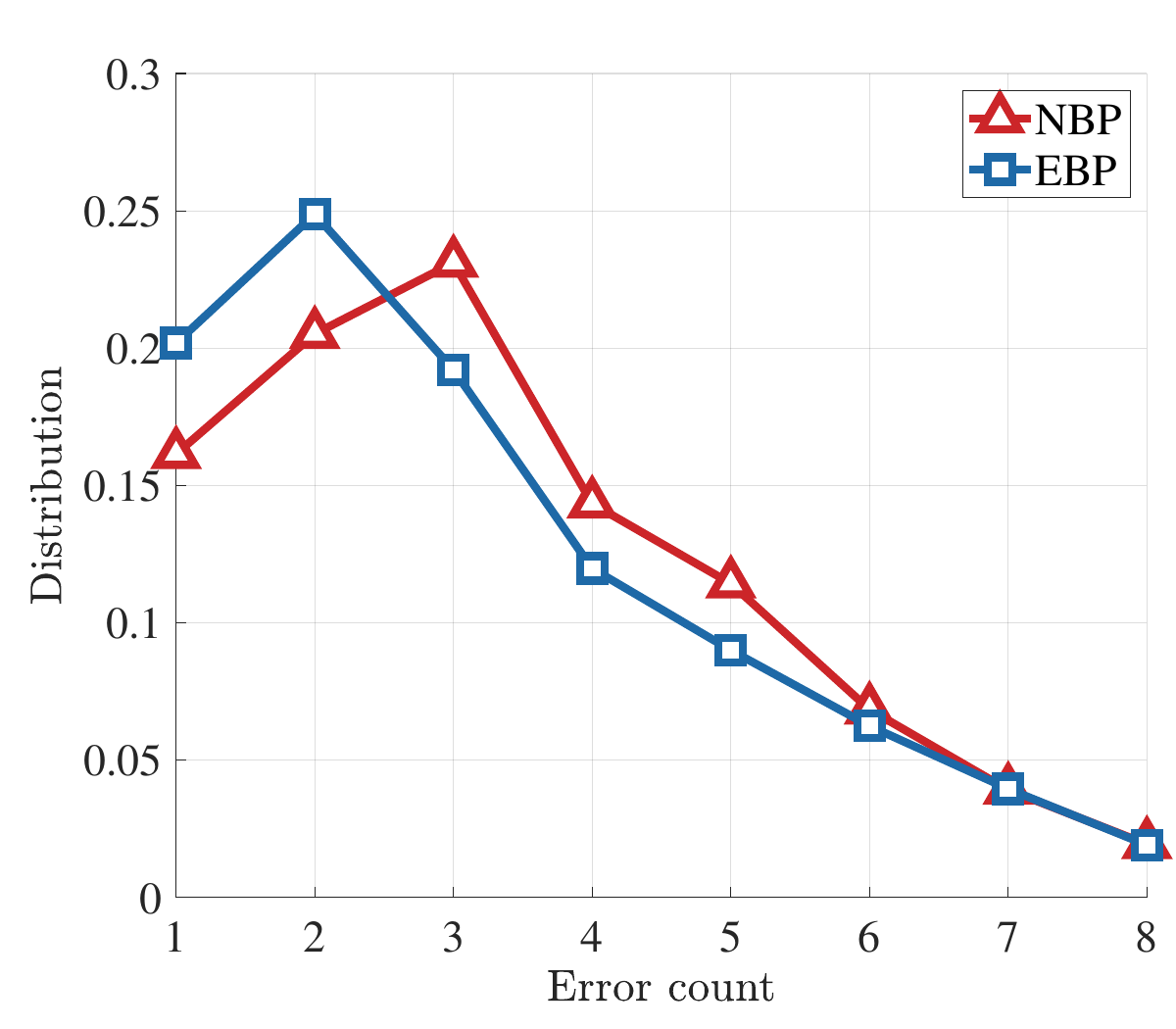}
\caption{Empirical probability mass functions (PMFs) of the residual error $\mathcal{E}\hat{\mathcal{E}}$ given a flagged failure of the pre-decoders (NBP and EBP) for the $d=7$ surface code at $p=0.05$.}
\label{Fig:Residual}
\end{figure}

\subsection{Performance Analysis}
Table~\ref{Table:Analysis} compares the performance metrics of various decoders for the $d=7$ surface code at $p=0.05$ and $\overline{\ell}=5$. The proposed sharing method enables the DE algorithm to find a weight set with a lower $\text{LER}_{+}$, at the expense of a slight increase in $\text{UFR}_{\text{P}}$. This trade-off is not problematic, since the magnitude of the $\text{UFR}_{\text{P}}$ increase is insignificant relative to the improvement in $\text{LER}_{+}$. Although the full parameterization contains the shared-weight configuration as a special case, weight sharing achieves better performance because it shrinks the search space, allowing the DE algorithm to find good solutions more easily, while still including weight configurations that are sufficient to break the message symmetry. In particular, we have found that extending the optimization of the full parameterization to $G=500$ generations provides only limited additional improvement, with its performance remaining below that of the shared-weight model.

In addition, the multi-objective selection rule (Multi.) effectively reduces $\text{FFR}_{\text{P}}$ at the cost of $\text{LER}_{+}$. The adaptive mutation strategy (Adap.) further accelerates convergence and leads to improved decoding performance in terms of both $\text{FFR}_{\text{P}}$ and $\text{LER}_{+}$. Consequently, we adopt weight sharing, multi-objective optimization, and adaptive mutation as the default configuration of the EBP+OSD decoder.

A comparison with baseline decoders focusing on the last three rows in Table \ref{Table:Analysis} provides several insights. In BP+OSD, decoding failures of the pre-decoder are predominantly flagged ($\text{UFR}_{\text{P}} \ll \text{FFR}_{\text{P}}$) and most of these flagged failures are successfully corrected by the OSD decoder, resulting in a competitive $\text{LER}{+}$. While NBP reduces $\text{FFR}_{\text{P}}$, it suffers from an increase in $\text{UFR}_{\text{P}}$. Since unflagged failures bypass the post-decoder, NBP+OSD yields a worse $\text{LER}_{+}$ than BP+OSD, highlighting the lack of end-to-end optimization. In contrast, EBP+OSD achieves the lowest $\text{LER}_{+}$ by suppressing $\text{FFR}_{\text{P}}$ without inducing a sharp increase in $\text{UFR}_{\text{P}}$. Moreover, the lower $\text{UFR}_{\text{O}}$, compared to NBP, indicates that EBP produces higher-quality and OSD-friendly inputs. Fig. 
\ref{Fig:Residual} compares the weight distributions of the residual error $\mathcal{E}\hat{\mathcal{E}}$ when the pre-decoder (NBP or EBP) declares a flagged failure at $p=0.05$ and $\overline{\ell}=5$. Each curve is obtained from more than $7000$ flagged-failure samples. It shows that EBP leaves fewer residual errors, which in turn leads to a lower $\text{UFR}_{\text{O}}$. This result demonstrates that the EBP decoder is explicitly optimized with the OSD post-decoder in mind. Finally, compared with BP+OSD in Table \ref{Table:Analysis}, EBP+OSD not only achieves a lower $\text{LER}{+}$ but also significantly reduces $\text{FFR}_{\text{P}}$, thereby decreasing the frequency of OSD activation and the overall computational complexity.

\begin{figure*}[t]
\begin{center}
\subfigure[]{\includegraphics[scale=0.35]{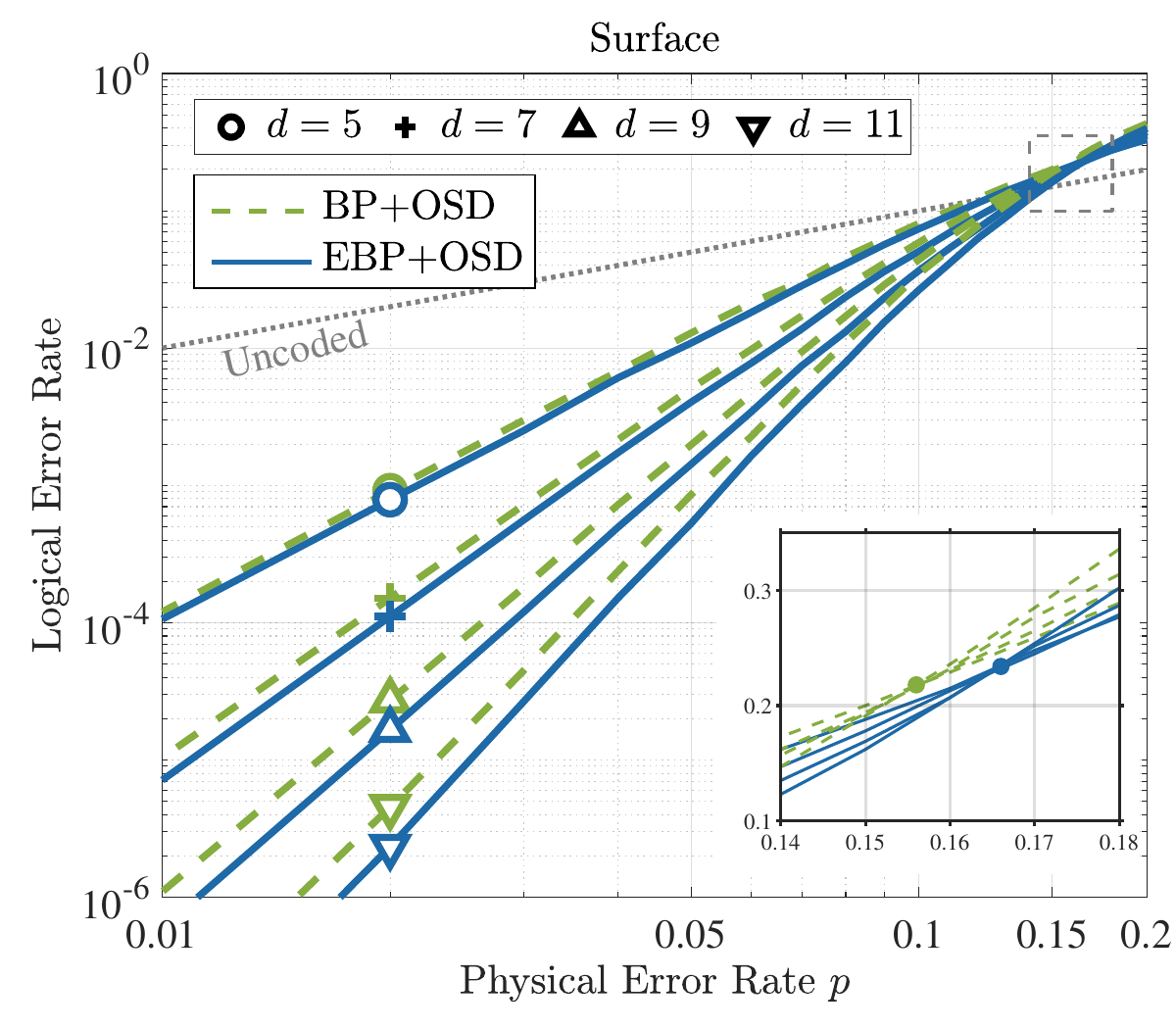}}
\quad
\subfigure[]{\includegraphics[scale=0.35]{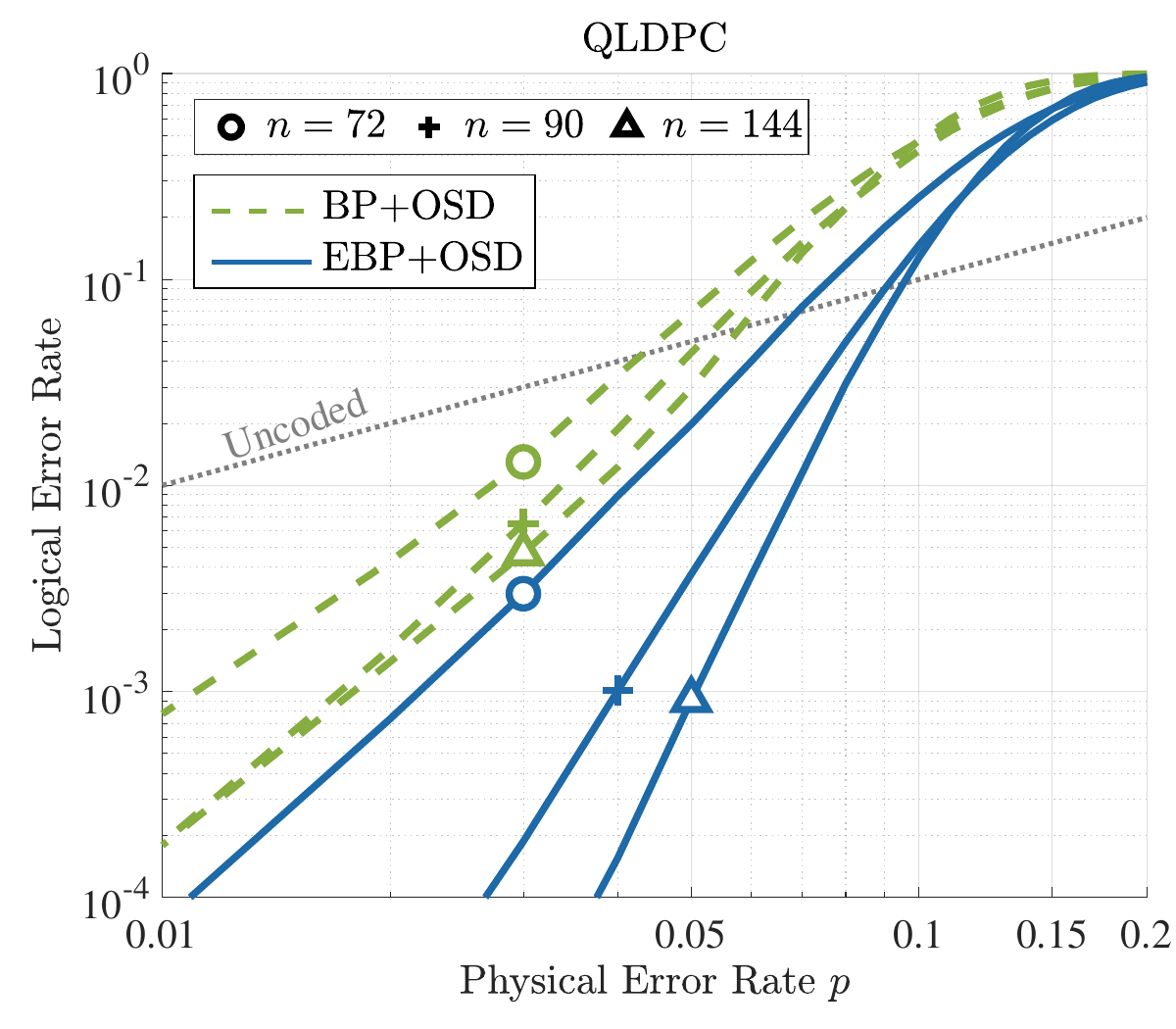}}
\subfigure[]{\includegraphics[scale=0.35]{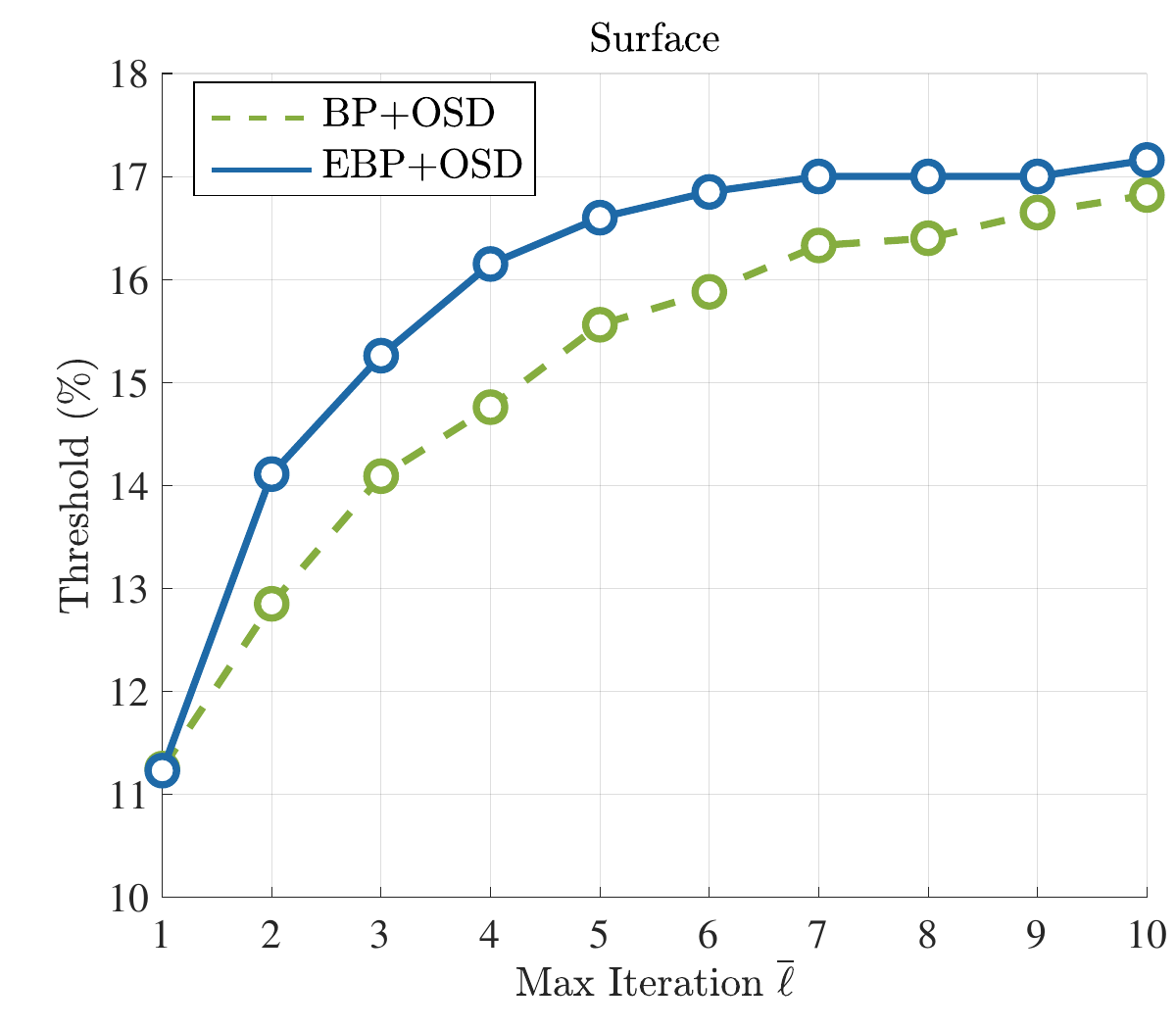}}
\quad
\subfigure[]{\includegraphics[scale=0.35]{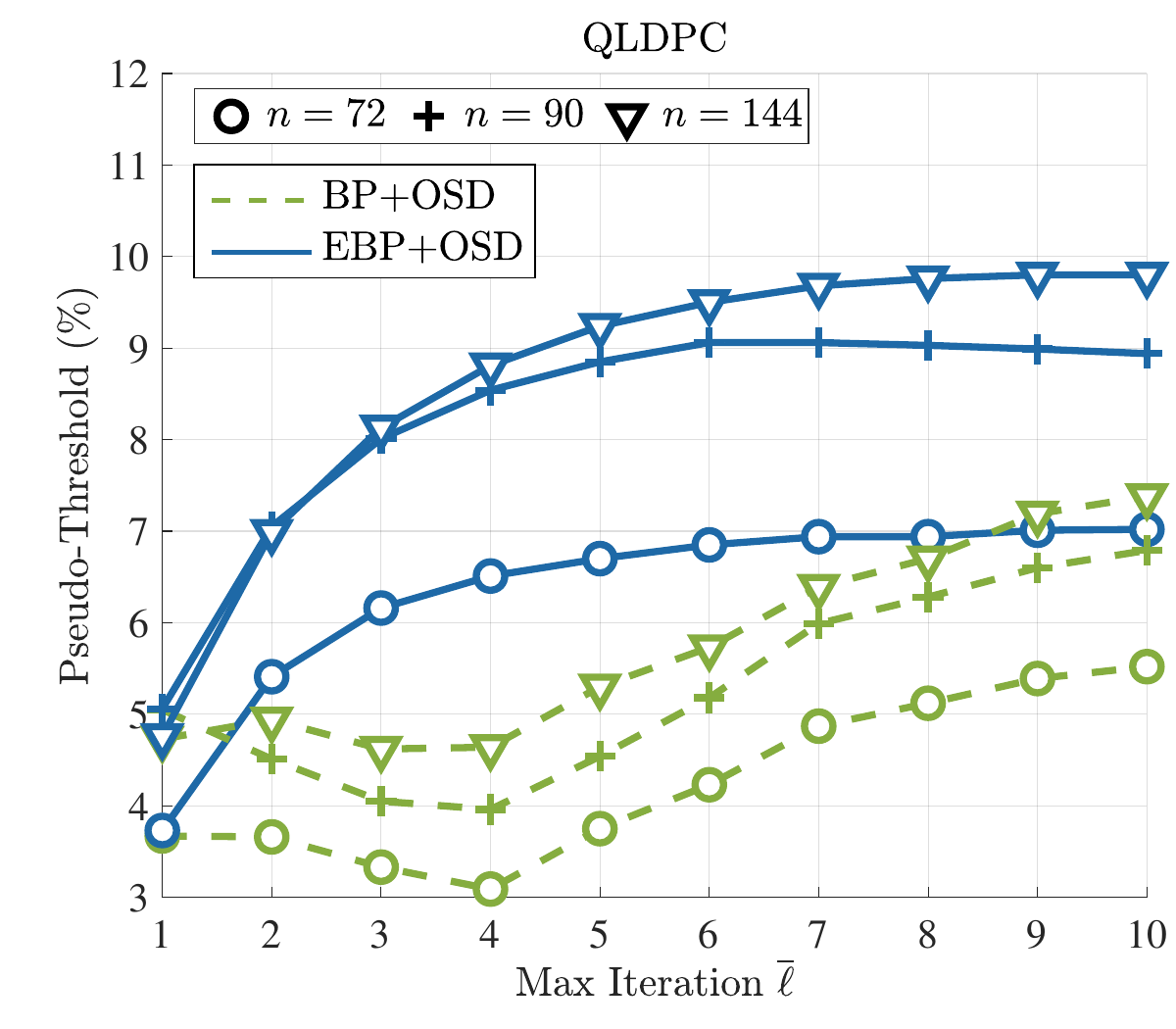}}
\end{center}
\caption{\label{fig:performance}Performance comparison between BP+OSD and EBP+OSD for surface codes (left column) and QLDPC codes (right column) in terms of LER (upper row) and threshold (lower row).}
\end{figure*}

\section{Simulation Results}
\label{sec_Results}
\subsection{Performance Comparison}
In this subsection, we evaluate the decoding performance using two representative code families: $[[d^2,1,d]]$ surface code~\cite{Fowler2012} and the bivariate bicycle (BB) QLDPC codes~\cite{Bravyi2024} with parameters $[[72,12,6]]$, $[[90,8,10]]$, and $[[144,12,12]]$.
Fig.~\ref{fig:performance} compares the performance of the BP+OSD and the EBP+OSD decoders for the surface codes (left column) and QLDPC codes (right column) in terms of the LER (upper row) and threshold (lower row). Each reported LER point is obtained by running at least $10^5$ Monte Carlo trials and continuing the simulation until at least $1000$ logical failures are collected, ensuring reliable statistics.

Figs.~\ref{fig:performance}(a) and (b) compare BP+OSD and EBP+OSD under a code-capacity depolarizing noise model with $\overline{\ell} = 5$. Using the proposed edge-indexed sharing method, we optimize the weight set on the surface code with $d = 7$ and subsequently reuse it for other surface codes. The optimized weights are given by
\begin{equation}
\setlength{\arraycolsep}{1.5pt}
\overline{w}^{(\ell)}=
\begin{bmatrix}
0.36 \\
0.67 \\
0.78 \\
1.06 \\
1.11 
\end{bmatrix},
w_r^{(\ell)}=
\begin{bmatrix}
2.13 & 2.38 & 1.09 & 1.10 & 0.93 & 1.08 \\
1.22 & 1.14 & 1.03 & 0.92 & 0.87 & 0.92 \\
1.52 & 1.33 & 1.22 & 1.07 & 1.05 & 1.20 \\
1.31 & 1.04 & 1.04 & 0.84 & 0.86 & 1.10 \\
1.14 & 1.25 & 1.17 & 0.81 & 0.96 & 1.25
\end{bmatrix}
\end{equation}
As shown in Fig.~\ref{fig:performance}(a), the EBP+OSD decoder achieves consistent LER improvement across all code distances $d$ for surface codes, increasing the threshold from $15.6\%$ to $16.6\%$. Threshold is defined as the crossing point among LER curves for different distances. Fig.~\ref{fig:performance}(c) presents the threshold as a function of the maximum number of iterations $\overline{\ell}$. A substantial improvement is observed at low iterations, demonstrating that EBP+OSD is particularly effective in low-latency decoding regimes.

\begin{table}[t]
\centering
\small
\caption{Threshold comparison of various decoders}
\label{table:threshold}
\begin{tabular}{l@{\hspace{0.4 em}}c@{\hspace{0.5em}}c@{\hspace{0.3em}}c@{\hspace{0.35em}}c@{\hspace{0.3em}}c}
\toprule
\multirow{3}{*}{Decoder} & \multirow{3}{*}{Iteration $\overline{\ell}$} & Threshold & \multicolumn{3}{c}{Pseudo-Threshold} \\
\cmidrule(lr){3-3} \cmidrule(lr){4-6}
& & Surface & Surface & \multicolumn{2}{c}{QLDPC} \\
& & & $d=7$ & $n=72$ & $n=144$ \\
\midrule
MWPM \cite{Edmonds1965} & -- & 14.7\% & 11.9\% & -- & -- \\
FFNN \cite{Varsamopoulos2017} & -- & -- & 12.4\% & -- & -- \\
CNN \cite{ChamberlandRonagh2018} & -- & -- & 11.9\% & -- & -- \\
CNN \cite{JungAliHa2024} & -- & -- & 13.2\% & -- & -- \\
%UF \cite{DelfosseNickerson2021} & -- & -- & -- & -- & -- \\
AMBP \cite{KuoLai2022} & $\ge150$ & 16.0\% & -- & -- & -- \\
\midrule
BP+OSD \cite{PanteleevKalachev2021} & 5 & 15.6\% & 12.9\% & 3.8\% & 5.3\% \\
\rowcolor{gray!10}
EBP+OSD & 5 & 16.6\% & 13.7\% & 6.7\% & 9.2\% \\
\rowcolor{gray!10}
EBP+OSD & 10 & 17.2\% & 13.8\% & 7.0\% & 9.8\% \\
\midrule
BP+TBD \cite{JungHa2025} & 20 & 15.8\% & 13.2\% & -- & -- \\
\rowcolor{gray!10}
EBP+TBD  & 6 & 16.0\% & 13.8\% & -- & -- \\
\midrule
BP+OSD$_4$ \cite{Kung2023} & 70 & 17.9\% & -- & -- & -- \\
\rowcolor{gray!10}
EBP+OSD$_4$  & 10 & 17.9\% & 14.0\% & -- & -- \\
\bottomrule

\end{tabular}
\end{table}

For QLDPC codes, the weights are optimized only for the $n = 72$ code and then reused for the $n=90$ and $n=144$ codes.
The optimized weights for QLDPC codes are given by
\begin{equation}
\setlength{\arraycolsep}{1.5pt}
\overline{w}^{(\ell)}=
    \begin{bmatrix}
0.16 \\
0.57  \\
0.94 \\
1.26  \\
1.48 
\end{bmatrix},
w_r^{(\ell)}=
\begin{bmatrix}
1.66 & 1.58 & 1.63 & 1.36 & 1.32 & 1.49 \\
1.07 & 1.03 & 1.02 & 0.98 & 0.93 & 1.00 \\
0.92 & 0.94 & 0.87 & 0.84 & 0.82 & 0.91 \\
0.88 & 0.89 & 0.84 & 0.82 & 0.83 & 0.89 \\
0.90 & 0.88 & 0.85 & 0.86 & 0.89 & 0.89
\end{bmatrix}
\end{equation}
As shown in Fig.~\ref{fig:performance}(b), the LER improvement for QLDPC codes is even more pronounced. The steeper LER decay of EBP+OSD results in a significant performance gap at low physical error rates.
Fig.~\ref{fig:performance}(d) shows that the pseudo-threshold, defined as the intersection between the LER curve and the physical error rate, improves by several percentage points at small values of $\overline{\ell}$. Notably, the BP+OSD decoder does not exhibit a monotonic increase in threshold as the maximum number of iterations $\overline{\ell}$ increases. In other words, simply strengthening the BP pre-decoder by allowing more iterations does not necessarily improve the overall BP+OSD performance. This highlights that optimizing the pre-decoder in isolation is insufficient for enhancing the performance of the combined pre+OSD structure.
In contrast, the EBP+OSD decoder demonstrates an almost monotonic threshold improvement because its weights are explicitly optimized to maximize the end-to-end EBP+OSD performance.

Table~\ref{table:threshold} compares the threshold performance of the EBP+OSD decoder with that of existing decoders.
For both surface and QLDPC codes, the EBP+OSD decoder with only 5 iterations outperforms minimum-weight perfect matching (MWPM) \cite{Edmonds1965}, machine learning–based decoders \cite{Varsamopoulos2017, ChamberlandRonagh2018, JungAliHa2024}, the conventional BP+OSD decoder \cite{PanteleevKalachev2021} with the same number of iterations, and even BP with additional memory effects (AMBP) \cite{KuoLai2022}, which requires a significantly larger number of iterations. While increasing iterations to $\overline{\ell}=10$ yields slight gains, the performance nearly saturates at this point as shown in Figs.~\ref{fig:performance}(c) and (d).

In addition, we demonstrate the flexibility of the proposed DE-based optimization by applying it to other post-processing techniques, such as the TBD \cite{JungHa2025} and the quaternary OSD (OSD$_4$) \cite{Kung2023}. TBD serves as a lightweight post-processing stage, whereas OSD$_4$ is a more complex yet powerful alternative. Notably, while the baseline BP+TBD requires 20 iterations to reach a threshold of $15.8\%$, the EBP+TBD decoder attains a superior threshold of $16.0\%$ using only 6 iterations. Similarly, EBP+OSD$_4$ with 10 iterations matches the threshold performance of BP+OSD$_4$ with 70 iterations. These results confirm that our performance-driven evolutionary optimization can be universally applied to enhance various pre+post decoding structures, regardless of their specific decoding operations.

\begin{figure}[t]
\subfigure[]{\includegraphics[scale=0.4]{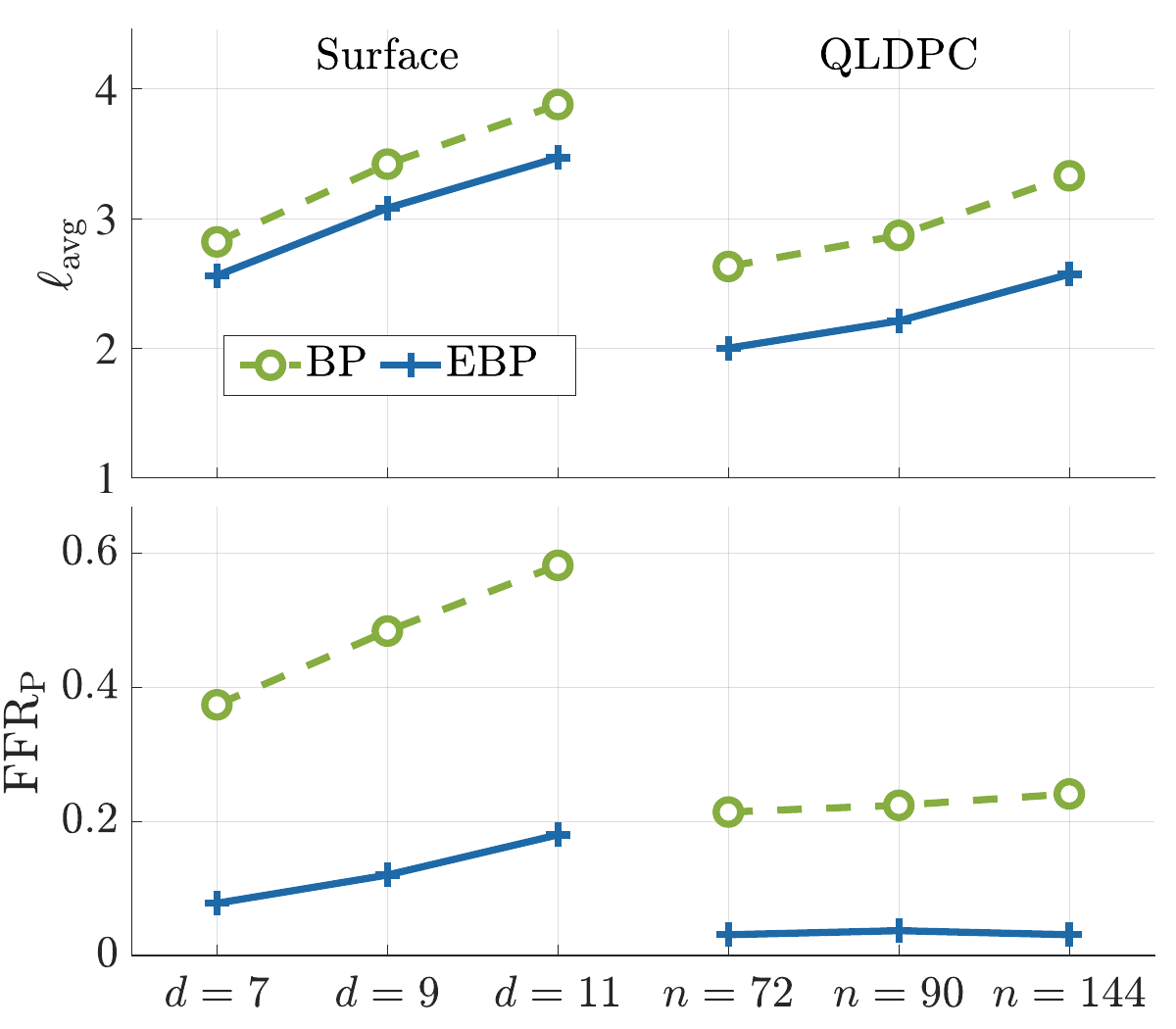}}
\subfigure[]{\includegraphics[scale=0.4]{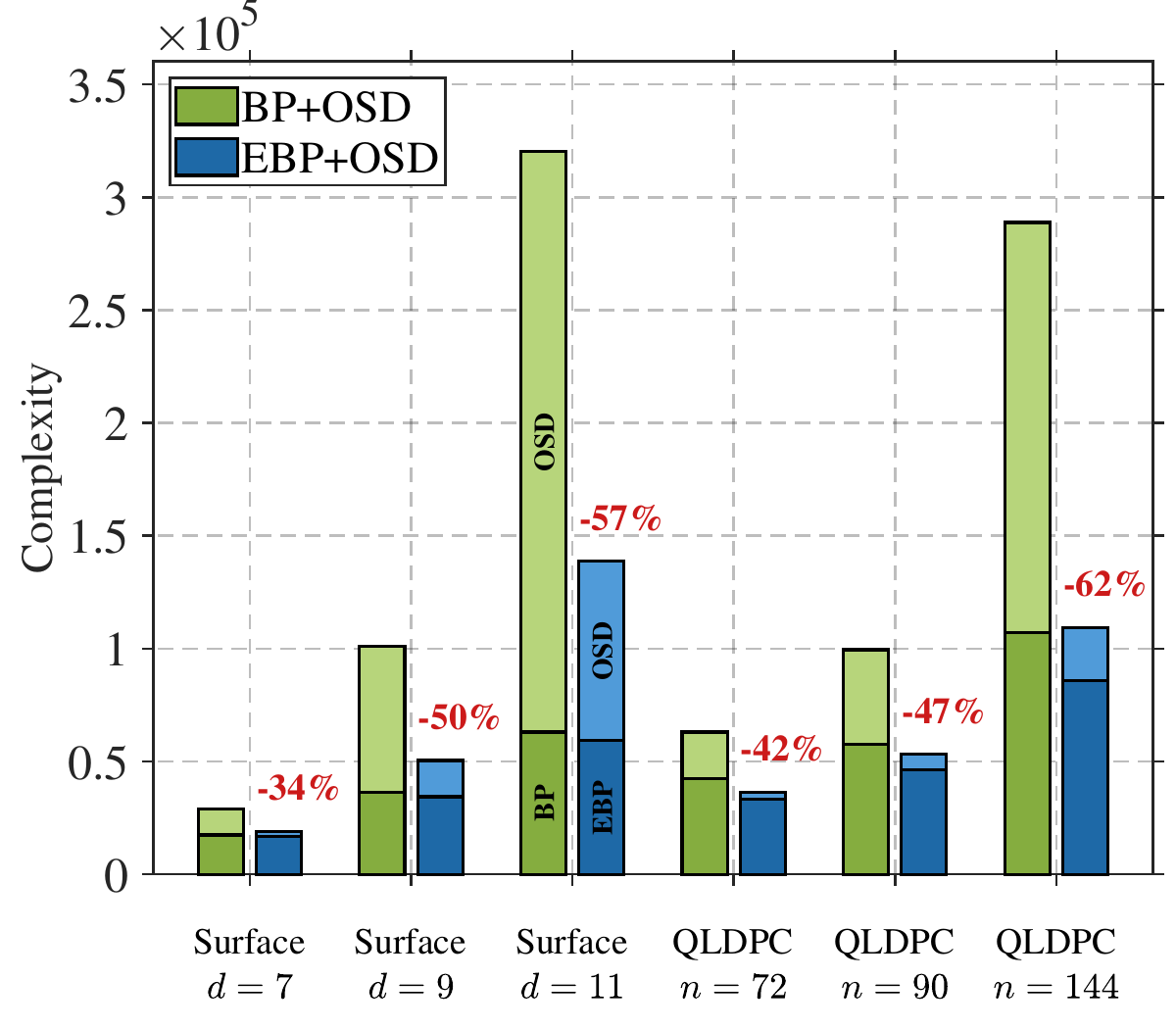}}
\caption{Comparison of decoding complexity between BP+OSD and EBP+OSD decoders: (a) EBP consistently reduces the average iteration count and ${\rm FFR}_{\rm P}$ for surface and QLDPC codes. (b) EBP+OSD achieves a 34–62\% reduction in total complexity across all codes.}
\label{fig:complexity}
\end{figure}

\subsection{Complexity Comparison}
In this subsection, we compare the computational complexity of the BP+OSD and EBP+OSD decoders. For an $[[n, k, d]]$ quantum code, $m_Z$ and $m_X$ are the numbers of $Z$-type and $X$-type CNs, respectively ($m = m_Z + m_X$). Similarly, let $e_Z$ and $e_X$ represent the numbers of edges connected to $Z$-type and $X$-type CNs, respectively ($e=e_Z+e_X$). Let $d_c$ denote the average CN degree.

The per-iteration complexity of the BP decoder is characterized as follows. First, the belief quantization step in Eq.~\eqref{Eq:Belief_Quantization} requires $4e$ lookup table (LUT) operations for exponential and logarithm functions. It also requires $2e$ additions and $e$ divisions. Second, the min-sum operation in Eq.~\eqref{Eq:Min} involves $2e$ XOR operations for sign computation, $(d_c + \lceil \log_2 d_c \rceil - 2)m$ comparisons to determine the minimum magnitudes, and $2e$ multiplications to combine the results. Third, the VN-side operations in Eqs.~\eqref{Eq:Weighted_Sum} and ~\eqref{Eq:Decision} require $2e_Z$, $2e_X$, and $2(e_Z+e_X)$ additions for the respective error types, totaling $4e$ additions. Finally, the EBP decoder introduces an additional overhead of $3n+e$ multiplications per iteration to apply weights.

The complexity of the OSD decoder is summarized as follows. Pre-computation of the probabilities $p_{j,\zeta}$ in Eq.~\eqref{Eq:OSD_p} requires $3n$ LUT operations, $3n$ additions for the denominator, and $3n$ divisions. For $X$-error processing, computing the unreliability measures $u_{j, X}$ in Eq.~\eqref{Eq:OSD_u} requires $n$ additions, and sorting incurs $n \log_2 n$ comparisons. Gaussian elimination applied to the permuted matrix and syndrome vector requires $m_Z(m_Z-1)(n+1)/2$ XOR operations for forward elimination and $m_Z(m_Z-1)/2$ XOR operations for backward substitution, resulting in the solution of the linear system. The same set of operations is performed for $Z$-error processing by replacing $m_Z$ with $m_X$.

To provide a quantitative comparison, we adopt a unit-cost model in which XOR and addition cost 1 unit, comparison costs 2 units, multiplication costs 1 unit, division costs 1 unit, and LUT operation costs 6 units \cite{Gamage2017}. The average complexity $\mathcal{O}_{+}$ of the pre+OSD decoder is modeled as:
\begin{equation}
\mathcal{O}_{+} = \mathcal{O}_{\text{P}} \times \ell_{\text{avg}} + \mathcal{O}_{\text{O}} \times \text{FFR}_{\text{P}},
\end{equation}
where $\mathcal{O}_{\text{P}}$ and $\mathcal{O}_{\text{O}}$ denote the per-iteration complexity of the pre-decoder and the complexity of the OSD decoder, respectively, and $\ell_{\text{avg}}$ is the average number of pre-decoder iterations.

Fig.~\ref{fig:complexity} compares the computational complexity $\mathcal{O}_{+}$ of the BP+OSD and EBP+OSD decoders at $p=0.05$ with $\overline{\ell}=5$. As shown in Fig.~\ref{fig:complexity}(a), both the average number of iterations $\ell_{\text{avg}}$ and the OSD activation probability $\text{FFR}_{\text{P}}$ are reduced for the EBP decoder compared with the BP decoder for both surface and QLDPC codes. Consequently, Fig.~\ref{fig:complexity}(b) shows that the EBP+OSD decoder achieves a substantial reduction in total complexity across all evaluated codes. This complexity reduction arises from two factors: the pre-decoder complexity is lowered due to the reduced $\ell_{\text{avg}}$, and the OSD complexity is mitigated by the decrease in $\text{FFR}_{\text{P}}$. The complexity reduction is even more pronounced for longer codes, where the OSD overhead is more dominant and the impact of reducing OSD activations is maximized. Overall, the total complexity reduction ranges from 34\% to as much as 62\%.

\begin{figure*}[t]
\centering
\subfigure[]{\includegraphics[scale=0.35]{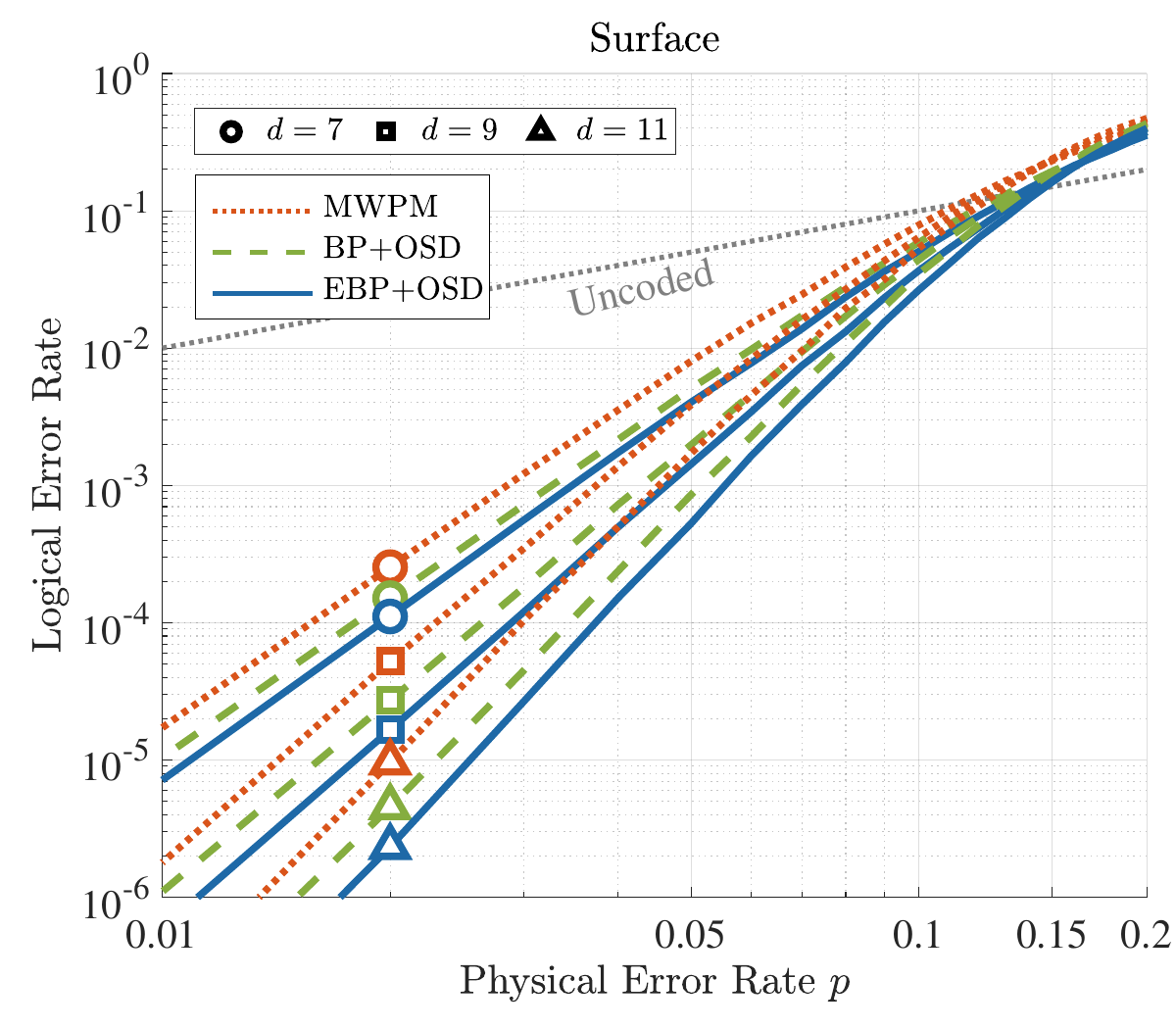}}
\subfigure[]{\includegraphics[scale=0.35]{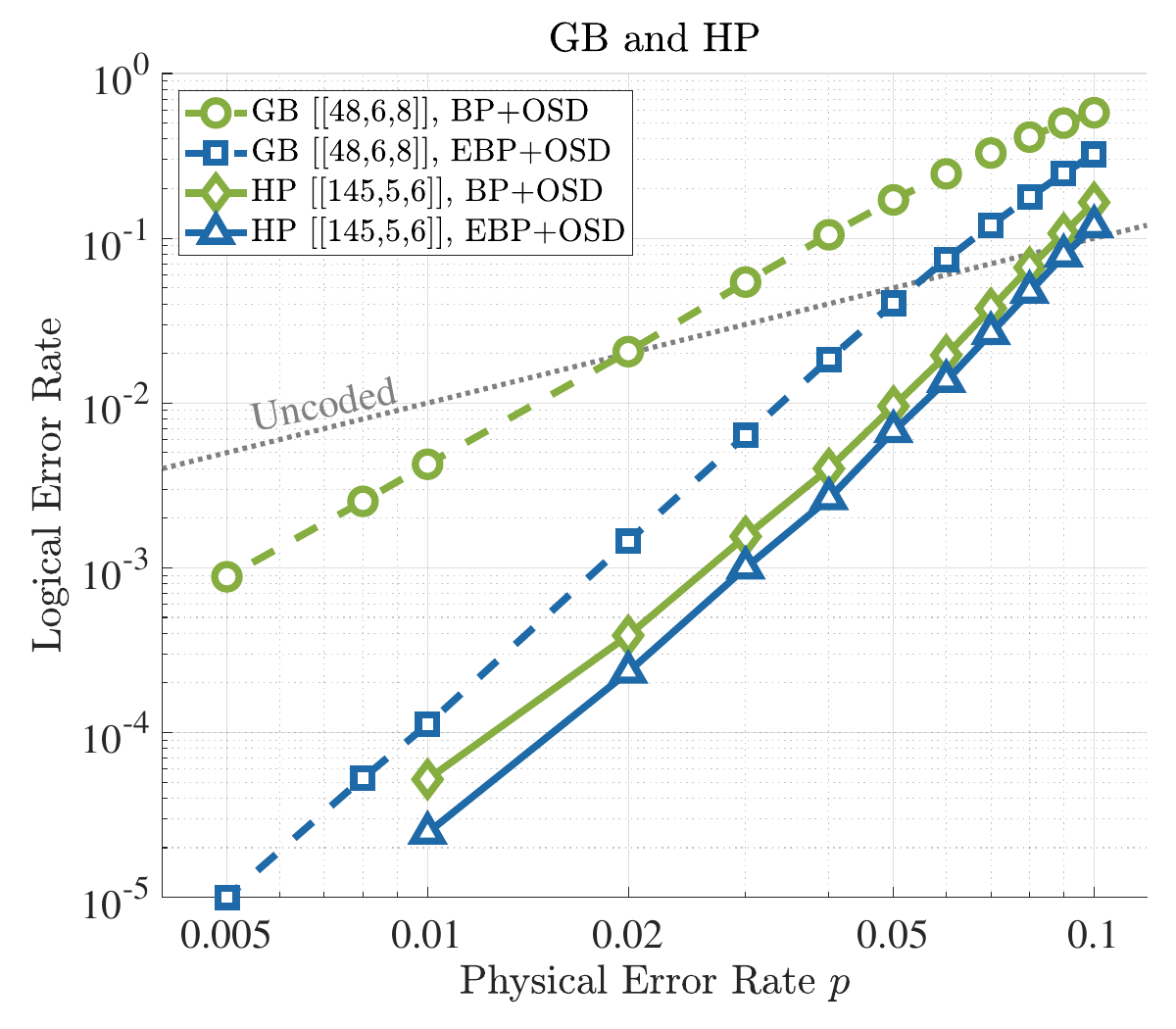}}
\subfigure[]{\includegraphics[scale=0.35]{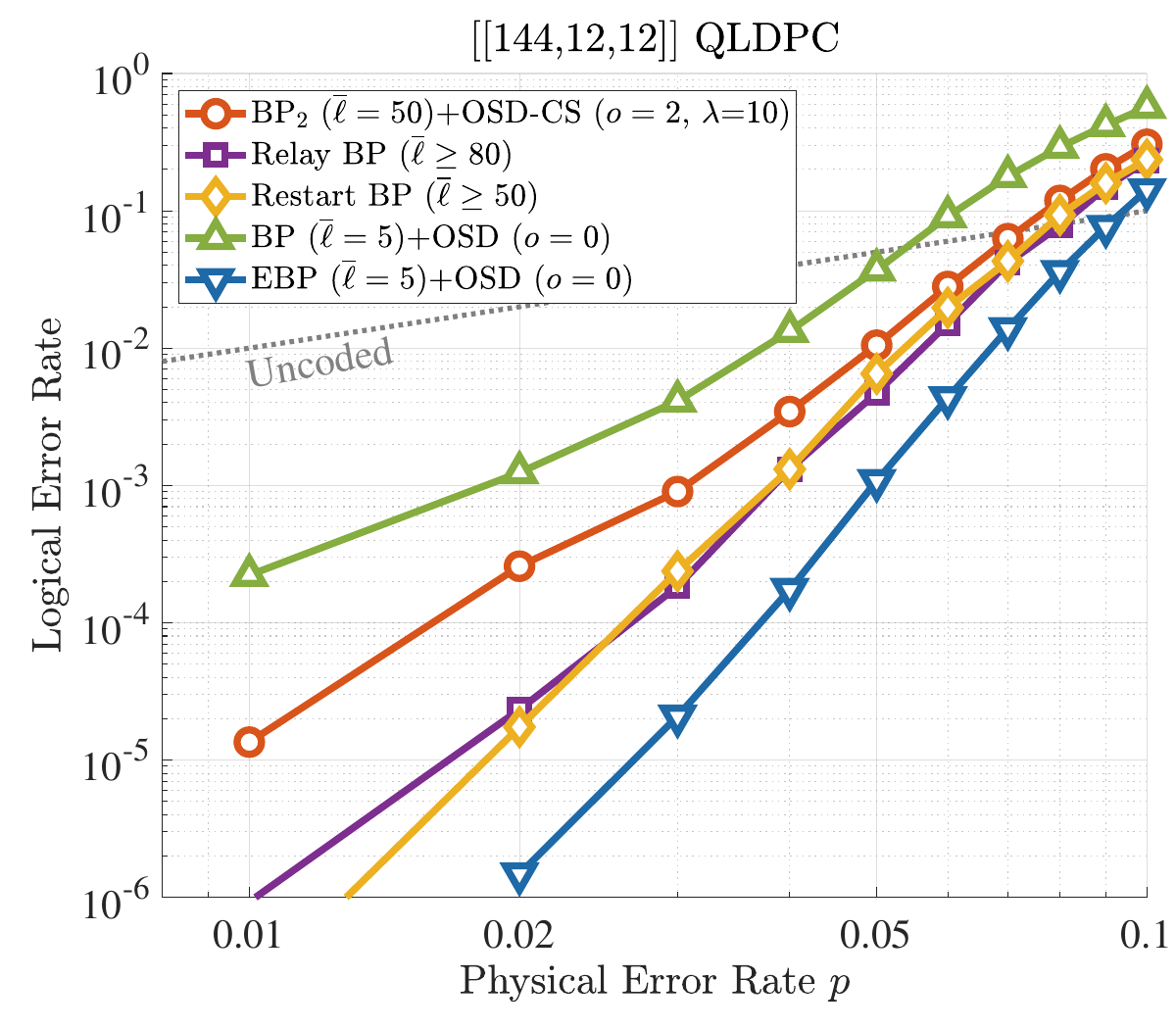}}
\subfigure[]{\includegraphics[scale=0.35]{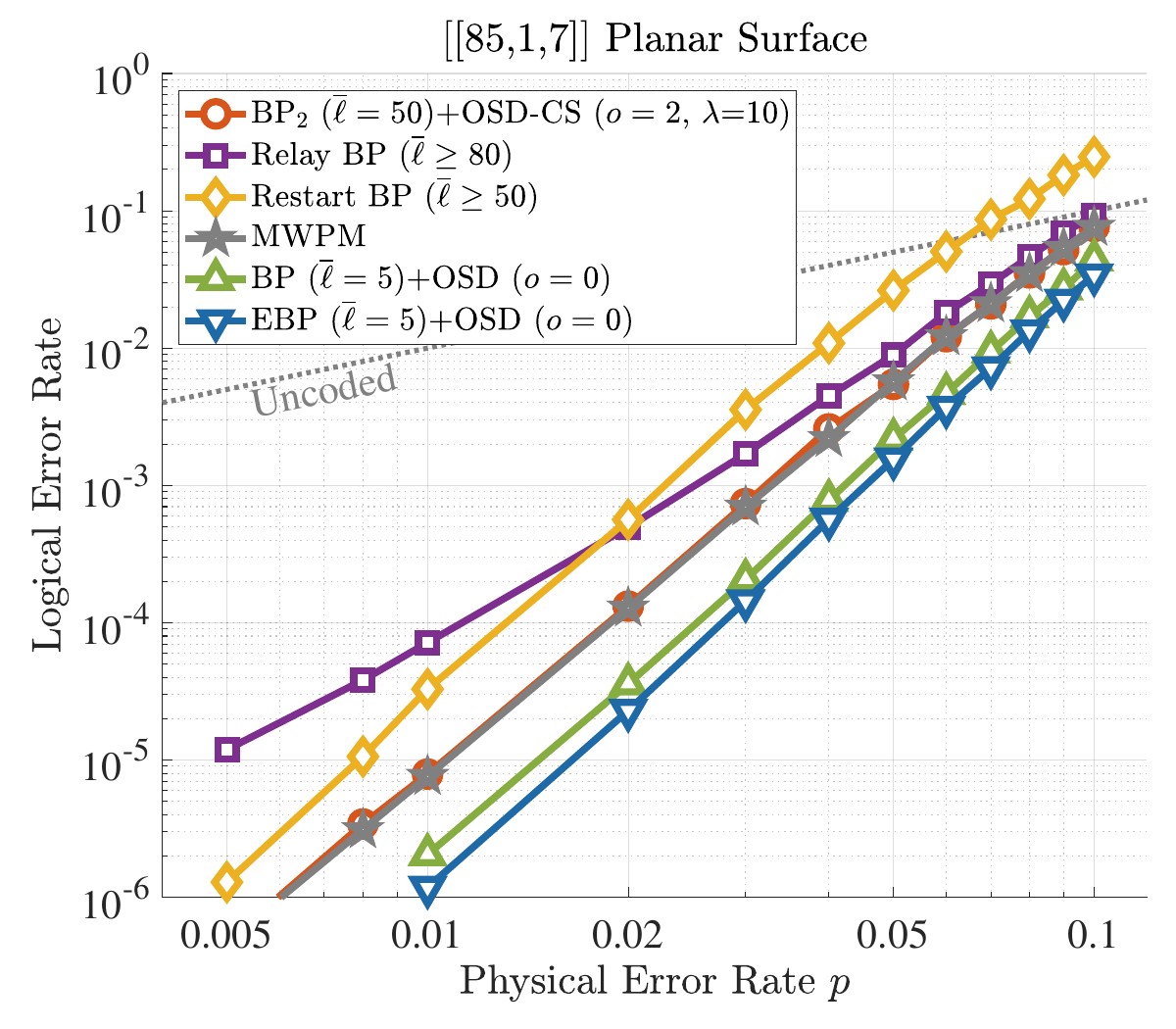}}

\caption{Additional benchmark comparisons. (a) LER comparison between MWPM, BP+OSD, and EBP+OSD for surface codes with distances $d=7,9,11$. (b) LER comparison for the $[[145,5,6]]$ hypergraph-product code and the $[[48,6,8]]$ generalized bicycle code. (c)--(d) LER comparison with relay BP, restart BP, and BP+OSD-CS for (c) the $[[144,12,12]]$ QLDPC code and (d) the $[[85,1,7]]$ planar surface code.}
\label{fig:add_benchmarks}
\end{figure*}

\begin{figure*}[t]
\centering
\subfigure[]{\includegraphics[scale=0.35]{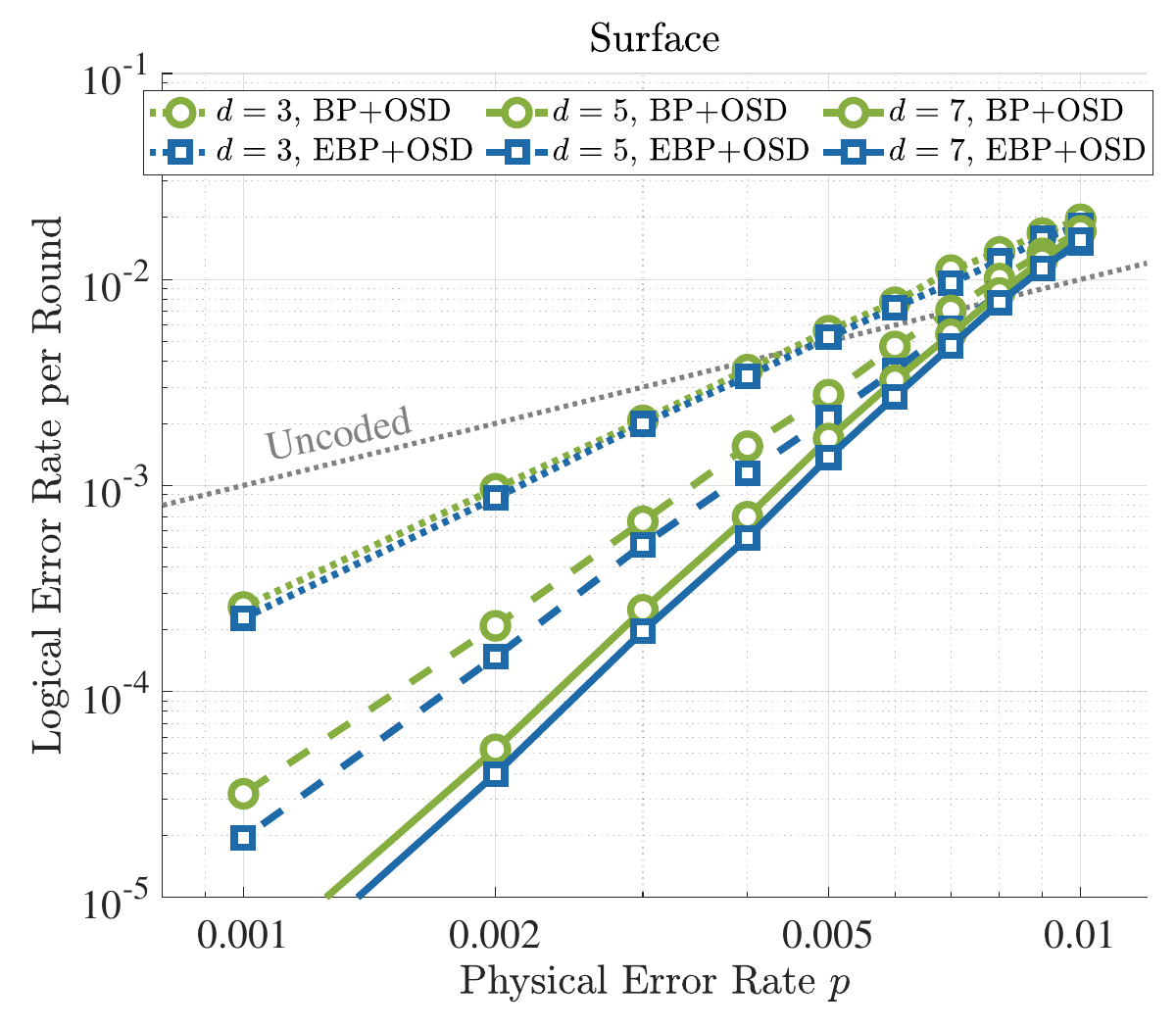}}
\subfigure[]{\includegraphics[scale=0.35]{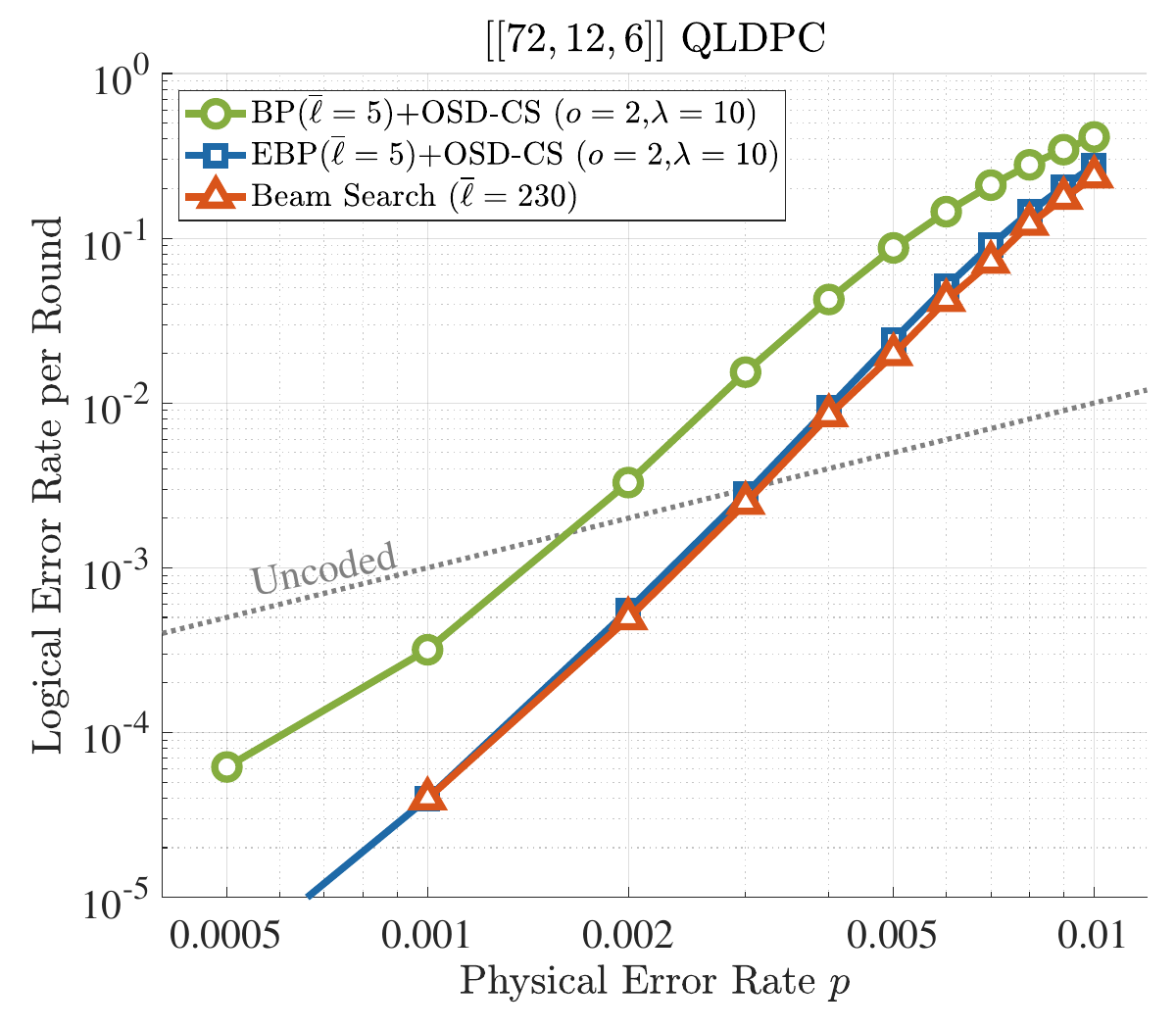}}
\caption{Circuit-level noise results, with LER normalized per syndrome-extraction round for the (a) surface codes and (b) $[[72,12,6]]$ QLDPC code.}
\label{fig:circuit_noise}
\end{figure*}

\subsection{Additional Decoder Benchmarks}
\label{app:additional_benchmarks}
This subsection provides additional benchmark comparisons, summarized in Fig.~\ref{fig:add_benchmarks}.
For surface codes, Fig.~\ref{fig:add_benchmarks}(a) presents a direct LER comparison with MWPM \cite{Edmonds1965} using a sparse-blossom implementation~\cite{HiggottGidney2023SparseBlossom}. The LER trend is consistent with the threshold comparison in Table~\ref{table:threshold}: EBP+OSD improves over both BP+OSD and MWPM.
To evaluate generalization across QLDPC constructions, Fig.~\ref{fig:add_benchmarks}(b) additionally considers a $[[145,5,6]]$ hypergraph-product (HP) code~\cite{RWBC2020} and a $[[48,6,8]]$ generalized bicycle (GB) code~\cite{KovalevPryadko2013}. The proposed EBP+OSD decoder consistently improves over BP+OSD for these QLDPC codes as well.

We further compare the proposed EBP+OSD decoder with advanced BP-based decoders. In Figs.~\ref{fig:add_benchmarks}(c) and (d), we compare EBP+OSD with relay BP~\cite{Muller2025Relay}, restart BP~\cite{Valentini2026}, and BP+OSD-CS~\cite{RWBC2020} for the $[[144,12,12]]$ QLDPC code~\cite{Bravyi2024} and the $[[85,1,7]]$ planar surface code~\cite{BravyiKitaev1998}, using the decoding parameters reported in each work. Relay BP uses $T_0=80$, $T_r=60$, $R=301$, and $S=5$, so that its worst-case iteration count $\overline{\ell}$ exceeds $80$ and grows with the number of relay legs. Restart BP uses $T_{\rm root}=50$, $T_{\rm branch}=10$, $\eta=35$, and $t=5$ for the $[[144,12,12]]$ code, giving a worst-case iteration count of at least $50$ that grows with the number of restarts; for the $[[85,1,7]]$ planar surface code, we use $T_{\rm root}=50$, $T_{\rm branch}=10$, $\eta=8$, and $t=3$. BP+OSD-CS uses binary BP with $\overline{\ell}=50$ iterations and the scaling factor $\alpha=1-2^{-\ell}$, followed by OSD-CS with order $o=2$ and $\lambda=10$. In contrast, the proposed EBP+OSD decoder uses quaternary BP with only $5$ iterations using the order-$0$ OSD post-decoder.
As shown in Figs.~\ref{fig:add_benchmarks}(c) and (d), EBP+OSD achieves the best LER among the compared methods for both the QLDPC code and the planar surface code. Although EBP+OSD uses OSD whereas restart BP and relay BP do not, it requires substantially fewer BP iterations. Compared with BP+OSD-CS, EBP+OSD also achieves strong performance using only a minimal BP iteration budget and an order-$0$ OSD.

\subsection{Evaluation over Circuit-Level Noise Model}
\label{app:circuit_noise}
The previous results consider the code-capacity noise model, which places errors only on the data qubits. In this subsection, we instead evaluate the proposed decoder under a circuit-level noise model, in which every operation in the syndrome-extraction circuit can introduce an error.
For surface codes, we perform a $Z$-memory experiment with $d$ syndrome-extraction rounds using the Stim stabilizer-circuit simulator~\cite{Gidney2021Stim}. For the $[[72,12,6]]$ QLDPC code, we adopt the circuit-level noise model provided in~\cite{Ye2026BeamSearch}. In both cases, the resulting detector error model (DEM) is converted into a binary parity-check matrix $\mathbf{H}_{\rm DEM}$, whose columns correspond to fault mechanisms and whose rows correspond to detectors, with $[\mathbf{H}_{\rm DEM}]_{i,j}=1$ when fault $j$ flips detector $i$. Decoding is performed by binary BP initialized with the fault probabilities provided by the DEM. Given the observed detector syndrome, BP produces posterior LLRs for the faults, which are then post-processed by OSD. As in the code-capacity case, EBP augments this pipeline with trainable weights; here, the CN weights are shared per detector, so that the weight set is $\{ \overline{w}^{(\ell)}, w_c^{(\ell)} \mid 1\le\ell\le \overline{\ell},\, 1\le c \le m\}$.

Fig.~\ref{fig:circuit_noise} compares the LER normalized per syndrome-extraction round for the surface codes and the QLDPC code. For the surface codes, EBP+OSD provides a consistent gain over BP+OSD, as in the code-capacity setting. For the QLDPC code, we adopt the stronger OSD-CS post-decoder (order $o=2$, $\lambda=10$); in this case, EBP+OSD achieves a substantial gain over BP+OSD and attains performance comparable to the state-of-the-art beam-search decoder~\cite{Ye2026BeamSearch}. Whereas the beam-search decoder uses up to $230$ BP-guided iterations, EBP+OSD instead relies on OSD and requires only $5$ BP iterations.

\section{Conclusion}
\label{sec_conclusion}
Although the BP+OSD decoder demonstrates strong performance across a wide range of quantum codes, it has not been explicitly optimized in an end-to-end manner and remains computationally expensive.
To address this limitation, we proposed the EBP decoder—an end-to-end optimizable pre-decoder trained using a DE algorithm.
The flexibility of DE enables multi-objective optimization that jointly accounts for decoding performance and computational complexity, even in the presence of the non-differentiable OSD decoder.
As a result, the EBP+OSD decoder achieves substantial improvement in both decoding performance and computational efficiency under strict low-latency constraints. 
For surface codes, it yields a 1\% threshold improvement and up to a 57\% reduction in complexity, while for QLDPC codes, it provides a 2.9–3.9\% threshold gain with up to a 62\% reduction in complexity. Moreover, whereas prior BP-based decoders typically require at least 20 iterations, EBP+OSD attains these improvements with only 5 BP iterations, thereby satisfying the strict low-latency constraints of practical QEC systems.
To the best of our knowledge, this is the first work to enhance BP decoding through evolutionary optimization, underscoring its potential to advance practical, low-latency QEC decoders.


\bibliographystyle{IEEEtran}
\bibliography{Reference}

\appendices

\begin{figure}[h]
\centering
\includegraphics[scale=0.35]{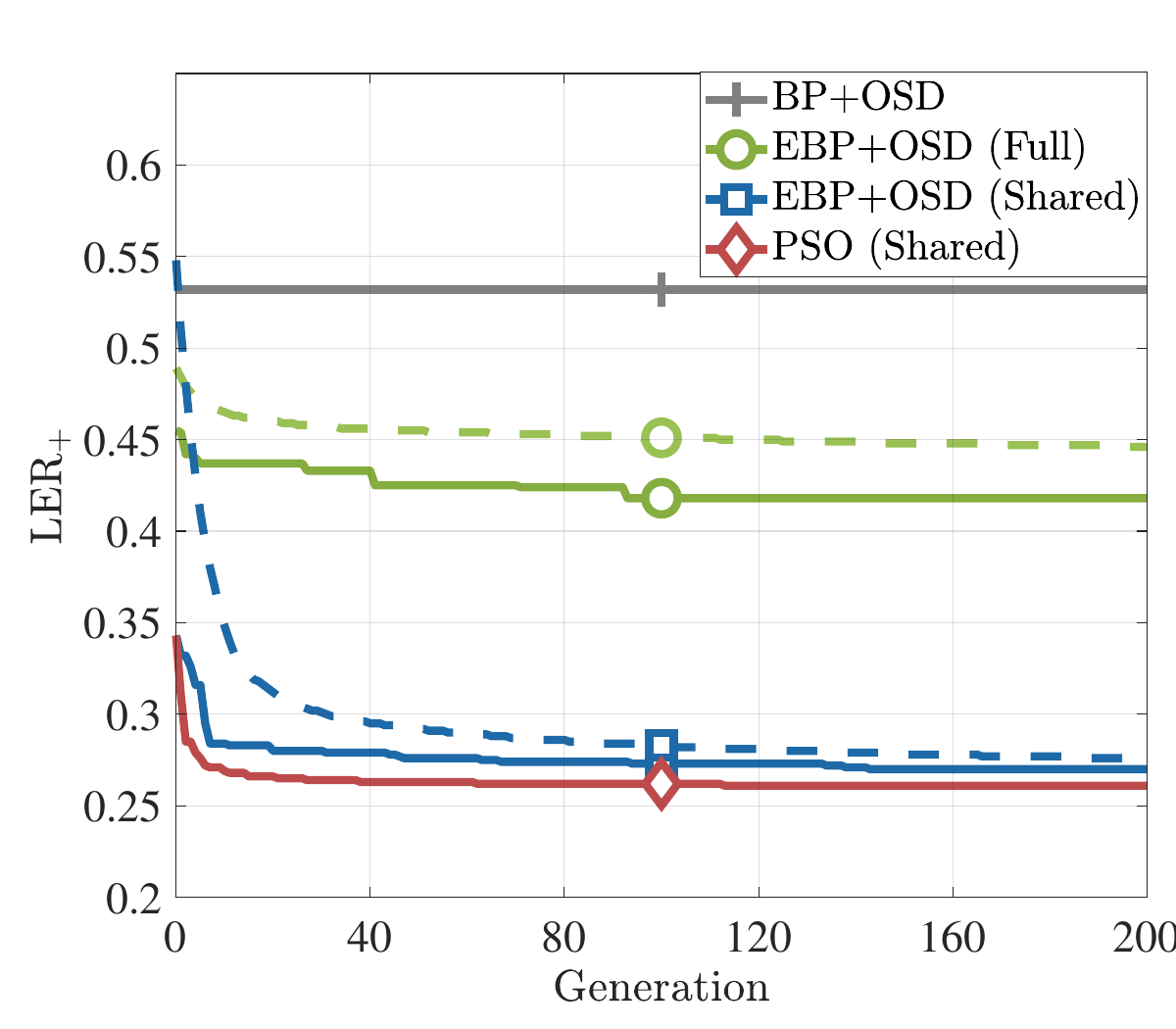}
\caption{Optimization convergence for the $[[72,12,6]]$ QLDPC code. As in the surface code, weight sharing is required for effective DE optimization, and replacing DE with PSO yields comparable performance.}
\label{Fig:Sharing_LER_QLDPC}
\end{figure}

\section{DE Convergence Analysis and PSO-Based Optimization for QLDPC Codes}
Besides DE, particle swarm optimization (PSO) is another widely used gradient-free optimizer. Fig.~\ref{Fig:Sharing_LER_QLDPC} shows the optimization results for the $[[72,12,6]]$ QLDPC code. As in Fig.~\ref{Fig:Sharing}(b) for surface codes, weight sharing yields better convergence for the QLDPC code as well. Moreover, although PSO converges faster in the early generations, it ultimately reaches performance comparable to DE. This indicates that the specific choice of gradient-free optimizer is not critical; what matters is the end-to-end optimization of BP together with the OSD post-decoder.

\begin{table}[h]
\centering
\small
\setlength{\tabcolsep}{2.5pt}
\caption{Performance analysis of EBP+OSD with various $\epsilon$ values.}
\label{Table:Epsilon_Analysis_Response}
\begin{tabular}{cccccc}
\toprule
\multirow{2}{*}{Decoder} & \multirow{2}{*}{$\epsilon$} & ${\rm UFR}_{\rm P}$ & ${\rm FFR}_{\rm P}$ & ${\rm UFR}_{\rm O}$ & ${\rm LER}_{\rm +}$ \\
 &  & $10^{-4}$ & $10^{-1}$ & $10^{-2}$ & $10^{-4}$ \\
\midrule
EBP+OSD & 0 & 0.22 & 2.35 & 2.69 & 0.86 \\
EBP+OSD & 0.01 & 0.25 & 1.85 & 3.04 & 0.81 \\
EBP+OSD & 0.05 & 0.37 & 1.12 & 4.38 & 0.86 \\
\bottomrule
\end{tabular}
\end{table}

\section{Effect of the Threshold Parameter $\epsilon$ in the Selection Rule}
Table~\ref{Table:Epsilon_Analysis_Response} reports the performance of EBP+OSD for the $d=7$ surface code under different values of the multi-objective threshold $\epsilon$. As $\epsilon$ increases, the selection rule places more emphasis on reducing ${\rm FFR}_{\rm P}$, which lowers the OSD activation probability. However, an excessively large $\epsilon$ sacrifices decoding performance and degrades ${\rm LER}_{+}$. Thus, $\epsilon$ is not a universal constant but a tunable parameter that can be adjusted according to the desired trade-off between decoding performance and OSD activation frequency.

\begin{figure}[h]
\centering
\includegraphics[scale=0.35]{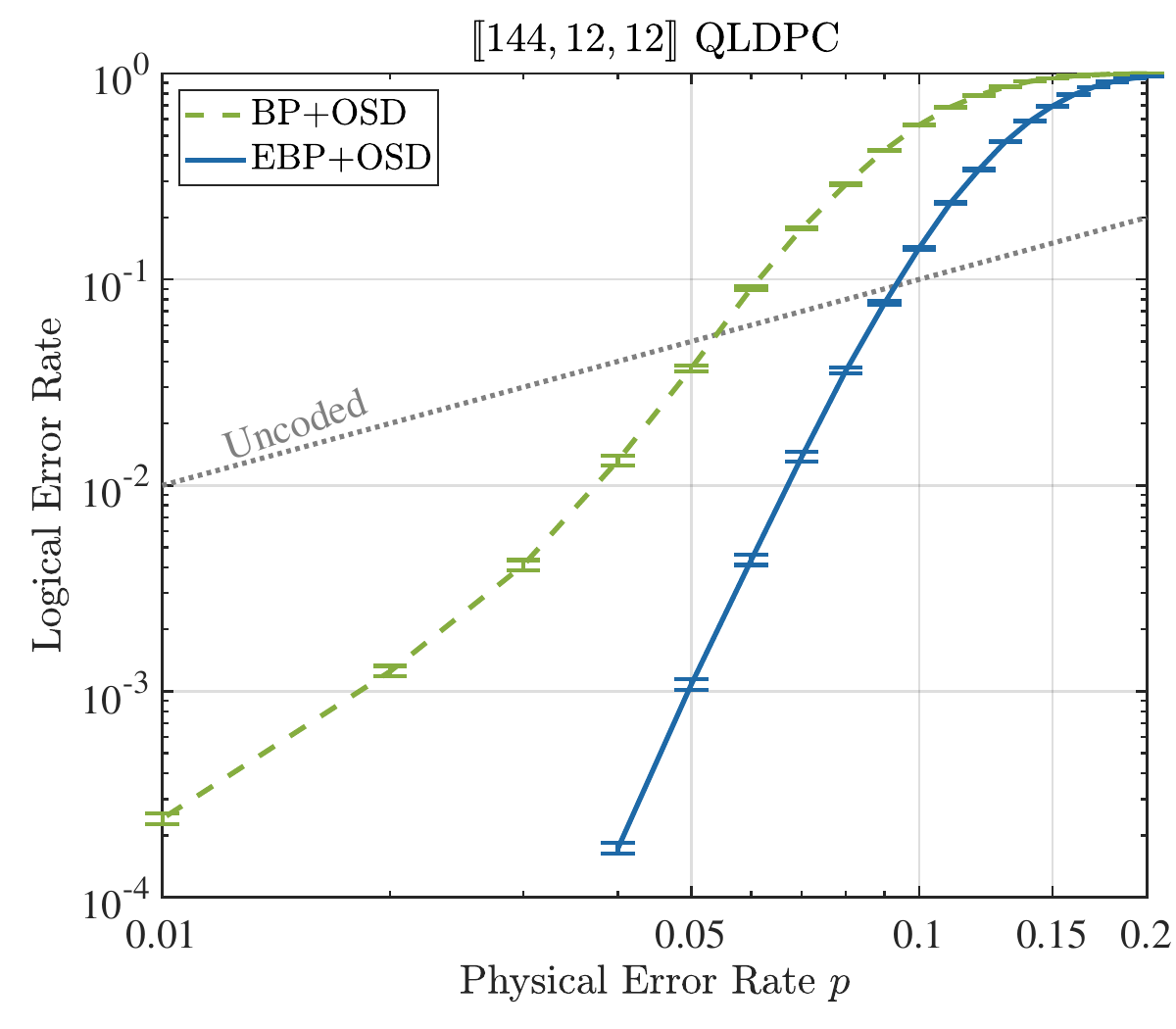}
\caption{LER of the $[[144,12,12]]$ QLDPC code, with error bars indicating $95\%$ Wilson binomial confidence intervals.}
\label{Fig:LER_QLDPC_error_bar}
\end{figure}

\section{Statistical Reliability of the Reported LERs}
As described in Section~\ref{sec_Results}, each reported LER point is obtained from at least $10^5$ Monte Carlo trials, with the simulation continued until at least $1000$ logical failures are collected, which yields statistically reliable estimates. Fig.~\ref{Fig:LER_QLDPC_error_bar} shows the LER of the $[[144,12,12]]$ QLDPC code together with error bars, where the error bars indicate $95\%$ Wilson binomial confidence intervals \cite{Brown2001}. As can be seen, the confidence intervals are very narrow, confirming that the reported LER values are statistically reliable.

\end{document}